\documentclass[twocolumn,aps,prl,notitlepage]{revtex4-2}
\pdfoutput=1
\usepackage{bm}
\usepackage{graphicx}
\usepackage{amssymb}
\usepackage{amsmath}
\usepackage{tensor}
\usepackage{dcolumn}
\usepackage{amsfonts}
\usepackage{upgreek}
\usepackage{comment}
\usepackage[margin=.7in]{geometry}
\usepackage[euler]{textgreek}
\usepackage{lineno}
\usepackage{braket}
\usepackage[caption=false]{subfig}
\usepackage{floatrow}
\usepackage{gensymb}
\usepackage{multirow}
\usepackage{wasysym}
\usepackage{mathtools}
\usepackage[usenames,dvipsnames]{xcolor}
\usepackage[normalem]{ulem}
\usepackage[colorlinks]{hyperref}
\hypersetup{
    colorlinks=true,
    citecolor=blue,
    linkcolor=blue,
    filecolor=magenta,      
    urlcolor=blue,
    }
\usepackage{lineno}
\usepackage{tikz} 
\allowdisplaybreaks
\newcommand{\stkout}[1]{\ifmmode\text{\sout{\ensuremath{#1}}}\else\sout{#1}\fi}
\newcommand{\bs}{\boldsymbol}
\newcommand{\pd}{\partial}
\newcommand{\pr}{{\prime}}
\newcommand{\dpr}{{\prime\prime}}

\newcommand{\mbk}{\mathbf{k}}

\newcommand{\veps}{\varepsilon}

\begin{document}

\title{Quantum response theory and momentum-space gravity}
\author{M. Mehraeen}
\email{mxm1289@case.edu}
\affiliation{Department of Physics, Case Western Reserve University, Cleveland, Ohio 44106, USA}
\date{\today}
\begin{abstract}
We present a quantum response approach to momentum-space gravity in dissipative multiband systems, which dresses both the quantum geometry--through an interband Weyl transformation--and the equations of motion. In addition to clarifying the roles of the contorsion and symplectic terms, we introduce the three-state quantum geometric tensor as a necessary element in the geometric classification of nonlinear responses and discuss the significance of the emergent terms from a gravitational point of view. We also identify a dual quantum geometric drag force in momentum space that provides an entropic source term for the multiband matrix of Einstein field equations.
\end{abstract}
\maketitle

\textit{Introduction.}$-$The space of physical states of a quantum mechanical system--complex projective Hilbert space--has the structure of a Kähler manifold~\cite{nakahara2018geometry}. This has inspired various geometric and topological interpretations of physical phenomena in quantum systems over the years~\cite{ashtekar1999geometrical, xiao2010berry}. And, in several areas of exploration, this quantum state geometry has recently garnered immense interest~\cite{torma2023essay, liu2024quantum, yu2024quantum, chen2024quantum, shim2025spin, jiang2025revealing, verma2025quantum}, with the identification of the role of the quantum metric~\cite{provost1980riemannian}--the real part of the quantum geometric tensor (QGT)--in an increasing number of physical effects~\cite{neupert2013measuring, claassen2015position, ahn2020low, watanabe2021chiral, wang2021intrinsic, liu2021intrinsic, bhalla2022resonant, gao2023quantum, wang2023quantum, hetenyi2023fluctuations, das2023intrinsic, komissarov2024quantum, onishi2024fundamental, fang2024quantum, kang2024measurements, jankowski2025optical}.

While the dynamical significance of the Berry curvature has been known for decades~\cite{karplus1954hall, kohn1957quantum, adams1959energy, chang1995berry, chang1996berry, sundaram1999wave}, it is perhaps the emergence of the quantum-metric Levi-Civita connection~\cite{gao2014field} that truly reveals the geometric underpinning of carrier dynamics and highlights the geodesic nature of the motion of Bloch electrons. Combining this with the Einstein field equations (EFE) arising from the Riemannian structure of quantum state manifolds has led to the intriguing recent proposal of momentum-space gravity~\cite{smith2022momentum} in condensed matter systems.

In this framework, the quantum metric in momentum space is viewed as the dual of the classical spacetime metric and the intrinsic dynamics is described by a dual Lorentz force. Furthermore, for mixed states, the momentum-space EFE acquire a source term that depends on the von Neumann entropy, which reveals intriguing similarities with efforts in understanding the fundamental nature of gravity and spacetime~\cite{bekenstein1973black, hawking1975particle, ruppeiner1979thermodynamics, jacobson1995thermodynamics, padmanabhan2010thermodynamical, verlinde2011origin, carroll2016what, bianconi2025gravity} and suggests a materials setting for exploring the connections between gravity and thermodynamics. Given the semiclassical nature of this approach based on wavepacket dynamics, a natural question that arises is the form this theory of gravity takes in momentum space from the vantage point of quantum response theory.

\begin{figure}[t]
\captionsetup[subfigure]{labelformat=empty}
    \sidesubfloat[]{\includegraphics[width=0.85\linewidth,trim={1cm 0cm 1cm 1cm}]{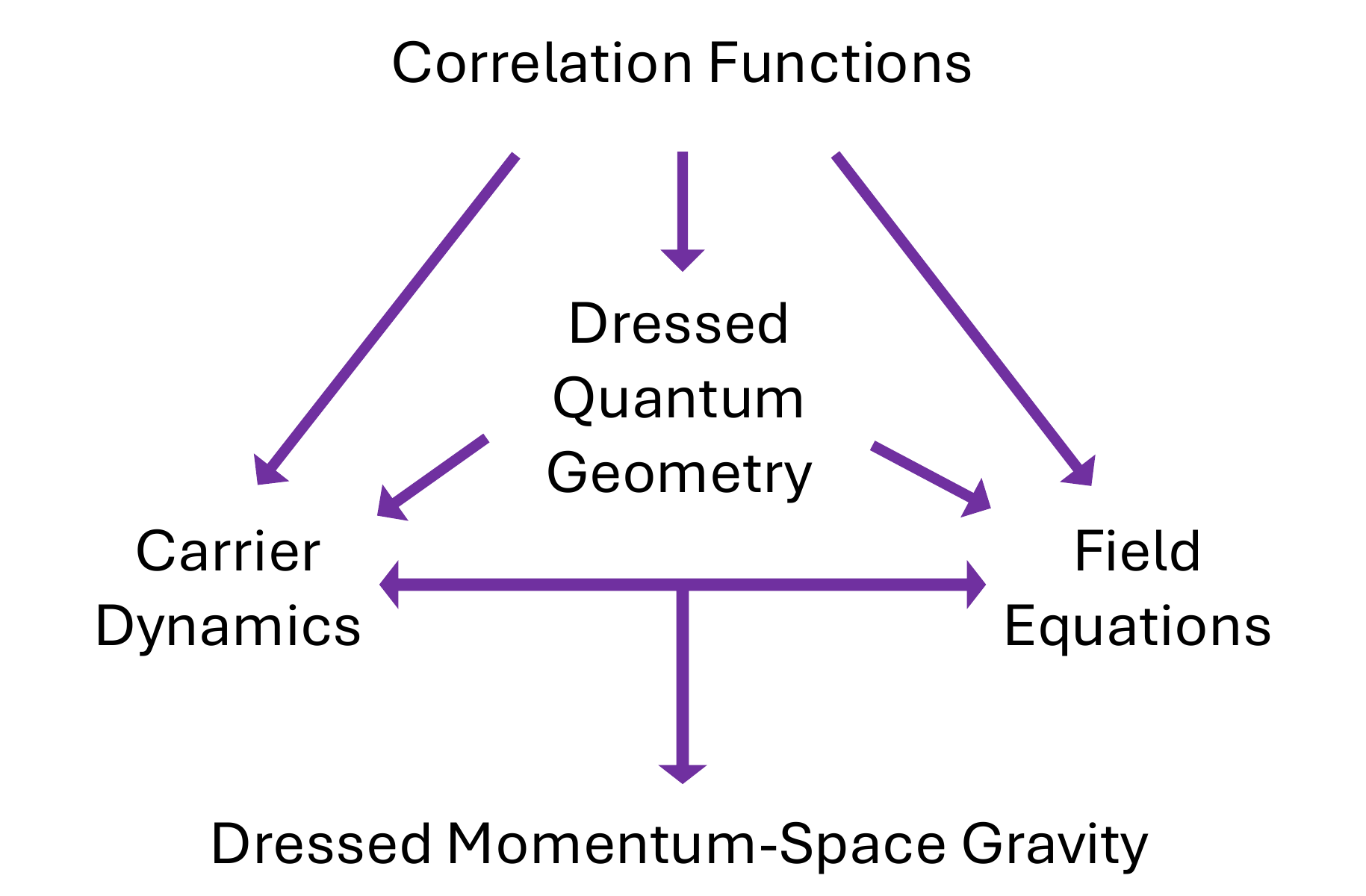}\label{fig1a}}
    \caption{Schematic summary of this work. Correlation functions corresponding to conductivity Kubo formulas are applied to simultaneously dress with dissipation the quantum geometry, carrier dynamics and momentum-space field equations. The latter two yield dressed versions of the two central equations of general relativity, resulting in dressed gravity in momentum space.}
    \label{fig1}
\end{figure}

Here, we address this question through a diagrammatic approach, with the general idea underlying this work presented schematically in Fig.~\ref{fig1}. We first apply Kubo formulas to propose a diagrammatic generalization of carrier dynamics to multiband systems in the presence of finite dissipation.  To this end, we build on recent developments~\cite{ahn2022riemannian, bouhon2023quantum, mitscherling2024gauge, avdoshkin2024multi, jankowski2024quantized, jankowski2025enhancing} in two-state~\footnote{Here, we refer to an $n$-state quantum geometric object as one that carries $n$ band indices. For example, $g_{\mu\nu}^a$ and $g_{\mu\nu}^{ab}$ are the single- and two-state quantum metric tensors, respectively.} quantum geometry to propose the \textit{three-state} QGT and show that it appears already at the quadratic-response level. We also discuss the appearance of the quantum geometric \textit{contorsion} tensor~\cite{nakahara2018geometry} and analogous symplectic terms when the full Berry covariant derivative is used to derive the Hermitian connection components. 

Physically, much intuitive insight has been gained from viewing the Berry curvature as an intrinsic gauge field strength tensor in momentum space~\cite{xiao2010berry}. Following this, and the standard argument in gravitational physics~\cite{misner1973gravitation, wald2010general}, we argue that once the choice of connection is made to be Levi-Civita, the remaining terms from the Hermitian connection and three-state QGT in the carrier position's equation of motion can then be interpreted as intrinsic matter fields arising from the multistate quantum geometry.

A noteworthy feature of the present approach is the appearance of dressed quantum geometric quantities in the equations of motion and EFE. While we discuss the gravitational significance of this later on, here we mention that there is increasing ongoing effort to define quantum geometry in interacting and disordered systems ~\cite{souza2000polarization, Michishita2022dissipation, chen2022measurement, kashihara2023quantum, zhou2024sloqvist, romeral2025scaling, sukhachov2025effect}. The approach which naturally emerges in the present derivation is closest to that of Ref.~\cite{Michishita2022dissipation} and, with the gravitational context in mind, is essentially a dissipation-induced multiband Weyl transformation of the quantum geometry. We apply this result to show that the dissipative correction to the EFE can be related to the entropic cost of dressing the metric through a dual momentum-space \textit{drag force} induced by the dissipative scattering. This, in turn, can be thought of as providing a diagrammatic generalization of the corresponding semiclassical arguments to dissipative multiband systems. We conclude by presenting an outlook on future insights that may be obtained from these results.

\textit{Multistate quantum geometry.}$-$In the presence of an applied electric field, the electric dipole interaction $e \mathbf{E} \cdot \mathbf{r}$ is included to the system Hamiltonian, where $\mathbf{r}$ is the position operator. In the k-space representation, this takes the form~\cite{blount1962formalisms}
$\mathbf{r}_{\mbk \mbk\pr}
=
-i \bs{\pd}_{\mbk\pr} \delta_{\mbk\pr \mbk}
+
\delta_{\mbk\pr \mbk} \bs{\mathcal{A}}_{\mbk}$, where $\mathcal{A}_{\mu \mbk}^{ab}
 =
 i \braket{u_{\mbk}^a | \pd_{\mu}^{\mbk} u_{\mbk}^b}$ is the Berry connection and $\ket{u_{\mbk}^a}$ the periodic part of the Bloch state~\footnote{Regarding notation, we use Greek letters for real- or momentum-space indices--with Einstein summation implied--and Latin letters for band indices. Furthermore, $\mathcal{O}_{(\mu\nu)} \equiv (\mathcal{O}_{\mu\nu} +  \mathcal{O}_{\nu\mu})/2$ and $\mathcal{O}_{[\mu\nu]} \equiv (\mathcal{O}_{\mu\nu} -  \mathcal{O}_{\nu\mu})/2$}. We henceforth work locally in momentum space and thus drop the momentum parameter from the notation. The commutator of the position operator and local momentum-space operators naturally defines a Berry covariant derivative
 $\bs{\mathcal{D}} \mathcal{O}
 \equiv
 -i [\mathbf{r},\mathcal{O}]$, which takes the form
\begin{equation}
(\mathcal{D}_{\mu} \mathcal{O})^{ab}
=
 \pd_{\mu} \mathcal{O}^{ab}
 -
 i \left[
 \mathcal{A}_{\mu}, \mathcal{O} \right]^{ab}.
\end{equation}
It should be stressed that this includes the full Berry connection, which is decomposed into diagonal and off-diagonal elements in band space as
$\mathcal{A}_{\mu}^{ab}
=
a_{\mu}^{a} \delta^{ab}
+
\mathcal{A}_{\mu}^{\pr ab}$.

The transition dipole matrix element $\mathcal{A}_{\mu}^{\pr ab}$ can be viewed as a component of the complex-valued vielbein matrix $e_{\mu}^{ab}
=
\mathcal{A}_{\mu}^{\pr ab} \ket{u^a} \bra{u^b}$ in band space~\cite{ahn2022riemannian}, thereby providing a geometric rationale for using off-diagonal Berry connection components as fundamental elements in the construction of multistate quantum geometric quantities. And the Hilbert-Schmidt inner product 
$\braket{A,B} = \text{Tr}(A^{\dagger} B)$
induces complex Riemannian structure on the manifold of quantum states and allows for the definition of quantum geometric invariants. The simplest such two-state object is the well-known QGT, which is obtained from the inner product of tangent basis vectors,
$Q_{\mu\nu}^{ab}
\equiv
\braket{e_{\nu}^{ab}, e_{\mu}^{ab}}
=
\mathcal{A}_{\mu}^{\pr ab} \mathcal{A}_{\nu}^{\pr ba}$, and which obeys the projector calculus identity
\begin{equation}
\label{eq_proj2}
i^2 \bra{u^a} P^a \pd_{\mu} P^b \pd_{\nu} P^a
\ket{u^a}
=
Q_{\mu\nu}^{ab}
-
\delta^{ab} \sum_c Q_{\mu\nu}^{ac},
\end{equation}
with the projection operator given by
$P^a = \ket{u^a}\bra{u^a}$. Note that for $b \neq a$, Eq.~(\ref{eq_proj2}) may be regarded as an alternative definition of the QGT in the projector calculus language. Indeed, this approach can be used to identify other elements of two-state quantum geometry~\cite{avdoshkin2024multi, mitscherling2024gauge}, with the two-state QGT being the primary member. The real and imaginary parts of this tensor define the two-state quantum metric and Berry curvature tensors,
$g_{\mu \nu}^{ab}
=
\text{Re} (Q_{\mu\nu}^{ab})$ and
$\Omega_{\mu \nu}^{ab}
=
-2 \text{Im} (Q_{\mu\nu}^{ab})$. Furthermore, we note that the two-state QGT is basically the band resolution of single-state quantum geometry, and yields the familiar Fubini-Study metric and (single-state) Berry curvature tensors once one sums over intermediate states, as is done in the second term on the right in Eq.~(\ref{eq_proj2}).

Moving beyond the QGT, the position operator acting on the tangent basis vectors induces a Hermitian connection~\cite{ahn2022riemannian}, the components of which are given by
$C_{\mu \nu \rho}^{ba}
=
\mathcal{A}_{\mu}^{\pr ab}
(\mathcal{D}_{\nu} \mathcal{A}_\rho^{\pr})^{ba}$. Using the standard definition of the (two-state) torsion tensor,
$\mathcal{T}_{\mu \nu \rho}^{ba}
=
2 C_{\mu [\nu \rho]}^{ba}$, we note that the real part of the Hermitian connection is expressed as
\begin{equation}
\text{Re} \left( C_{\mu \nu \rho}^{ba}\right)
=
\Gamma_{\mu \nu \rho}^{ba}
+
\text{Re} \left( K_{\mu \nu \rho}^{ba} \right),
\end{equation}
where
$\Gamma_{\nu \rho \mu}^{ba}
=
\frac{1}{2} 
( \pd_{\mu} g_{\nu \rho}^{ba}
+
\pd_{\rho} g_{\mu \nu}^{ba}
-
\pd_{\nu} g_{\mu \rho}^{ba})$
is a Levi-Civita connection component of the band-resolved quantum metric and the second term, given by
\begin{equation}
K_{\mu \nu \rho}^{ba}
=
\frac{1}{2}
\left(
\mathcal{T}_{\mu \nu \rho}^{ba}
-
\mathcal{T}_{\nu \mu \rho}^{ba}
-
\mathcal{T}_{\rho \mu \nu}^{ba}
\right),
\end{equation}
is identified as the quantum geometric contorsion tensor~\cite{nakahara2018geometry}, which contains the torsionful part of the metric connection and arises once one uses the full Berry covariant derivative to define the Hermitian connection, as opposed to using only the diagonal Berry connection components. We note that the contorsion tensor inherits the antisymmetric property
$K_{\mu \nu \rho}^{ba}
=
K_{[\mu \nu] \rho}^{ba}$ from the torsion tensor.

Motivated by a geometric characterization of nonlinear responses, we now present an extension of this framework beyond the two-state formalism by introducing the three-state QGT 
\begin{equation}
Q_{\mu \nu \rho}^{abc}
=
\mathcal{A}_{\mu}^{\pr ab} \mathcal{A}_{\nu}^{\pr bc} \mathcal{A}_{\rho}^{\pr ca},
\end{equation}
which cyclically connects three distinct states, and represents a natural generalization of its two-state counterpart. Formally, this can be defined by utilizing the non-Abelian QGT~\cite{provost1980riemannian, ma2010abelian, bouhon2023quantum, jankowski2025enhancing} in operator form, composed of basis vector derivatives, as
$\mathcal{Q}_{\rho \nu}^c
=\ket{\pd_{\rho} u^c} \bra{\pd_{\nu} u^c}$, which allows for the definition
\begin{equation}
Q_{\mu \nu \rho}^{abc}
\equiv
\Braket{\mathcal{Q}_{\rho \nu}^c, e_{\mu}^{ab}},
\end{equation}
with
$\mathcal{(Q}_{\nu \rho}^c)^{ba}
=
\mathcal{A}_{\nu}^{\pr bc} \mathcal{A}_{\rho}^{\pr ca}$. This naturally generalizes the two-state projector identity given by Eq.~(\ref{eq_proj2}) to the three-state identity
\begin{equation}
\label{eq_proj3}
\begin{split}
i^3 \bra{u^a} P^a \pd_{\mu} P^b \pd_{\nu} P^c \pd_{\rho} P^a
\ket{u^a}
=
Q_{\mu\nu\rho}^{abc}
-
\delta^{ab} \sum_d Q_{\mu\nu\rho}^{adc}
\\
+
\delta^{ab} \sum_d Q_{\mu\nu\rho}^{acd}
-
\delta^{bc} \sum_d Q_{\mu\nu\rho}^{abd},
\end{split}
\end{equation}
lending further support to the proposition that the three-state QGT defined above is indeed a central object in higher-state quantum geometry.

We point out that the real and imaginary parts of the three-state QGT are decomposed into linear combinations of the non-Abelian quantum metric and Berry curvature tensors as
\begin{subequations}
\begin{align}
\text{Re} \left( Q_{\mu\nu\rho}^{abc} \right)
&=
\mathcal{A}_{\mu}^{\pr (ab)}
(\textbf{g}_{\nu \rho}^c)^{ba}
-
\frac{i}{2}
\mathcal{A}_{\mu}^{\pr [ab]}
(\bs{\Omega}_{\nu \rho}^c)^{ba},
\\
\text{Im} \left( Q_{\mu\nu\rho}^{abc} \right)
&=
-i \mathcal{A}_{\mu}^{\pr [ab]}
(\textbf{g}_{\nu \rho}^c)^{ba}
-
\frac{1}{2}
\mathcal{A}_{\mu}^{\pr (ab)}
(\bs{\Omega}_{\nu \rho}^c)^{ba},
\end{align}
\end{subequations}
thereby exhibiting a more complex structure compared to the two-state counterpart. The symmetric and antisymmetric parts,
$S_{\mu\nu\rho}^{abc}
=
Q_{\mu(\nu\rho)}^{abc}$
and
$A_{\mu\nu\rho}^{abc}
=
Q_{\mu[\nu\rho]}^{abc}$, are also expressible in terms of these quantities. Specifically, we note that the antisymmetric term is the band resolution of the torsion tensor, i.e.,
$\sum_c A_{\mu\nu\rho}^{abc}
=
i \mathcal{T}_{\mu\nu\rho}^{ba}/2$, while 
$S_{\mu\nu\rho}^{abc}$ is its symmetric counterpart. As is shown in the next section, the three-state QGT already makes an appearance at the quadratic-response level in dissipative multiband systems and is required for a complete geometric characterization of the response functions.

\textit{Dressed carrier dynamics from diagrammatics.}$-$Motivated by recent developments via density-matrix methods in generalizing carrier dynamics~\cite{atencia2022semiclassical,  mehraeen2024quantum}, here, we obtain the equation of motion for the carrier position by taking the route of Kubo formulas. Using recently developed diagrammatic approaches within the Matsubara formalism~\cite{parker2019diagrammatic, Michishita2021effects}, the linear and quadratic ac  conductivities are obtained~\footnote{See the Supplemental Material at [URL] for details on the evaluation of the response functions, equations of motion, renormalization functions and the dissipative EFE. This includes Refs.~\cite{mahan2000many, carroll2019spacetime}}. In order to obtain dissipative carrier dynamics, we note that the dc current density, 
$j_{\mu}
= 
\sigma_{\mu\nu}E^{\nu}
+
\sigma_{\mu\nu\rho}E^{\nu} E^{\rho}$, is related to the equation of motion of the carrier position and the carrier density $n$ as
$j_{\mu}
=
-e \text{Tr}(n \dot{x}_{\mu})$~\cite{atencia2022semiclassical, mehraeen2024quantum}. Therefore, upon evaluating  the correlation functions, and using the equation of motion of the carrier momentum,
$\dot{k}^{\mu} = -e E^{\mu}/\hbar$,
we arrive at the equation of motion for the carrier position, which is expressed as
\begin{equation}
\label{eom}
\begin{split}
\dot{x}_{\mu}^a
&=
v_{\mu}^a
+
\dot{k}^{\nu} \sum_b 
(
\mathcal{Z}_{\Omega}^{ab} \tilde{\Omega}_{\mu \nu}^{ab}
-
\mathcal{Z}_{g}^{ab} \tilde{g}_{\mu \nu}^{ba}
+
\dot{k}^{\rho} m^{ab} \mathcal{Z}_{\Gamma}^{ab}
\tilde{\Gamma}_{\nu \rho \mu}^{ba})
\\
&+
\dot{k}^{\nu} \dot{k}^{\rho}
\sum_{b} 
( \mathcal{Z}_{m}^{ab} 
\tilde{g}_{\nu \rho}^{ba} \pd_{\mu}
+
\mathcal{M}_{\mu \nu \rho}^{ba}
+
\sum_c \mathcal{N}_{\mu \nu \rho}^{abc} ) m^{ab},
\end{split}
\end{equation}
where $v_{\mu}^a = \pd_{\mu} \veps^a / \hbar$ is the group velocity and the $\mathcal{Z}$'s are various renormalization functions that depend on the dissipation parameter. Eq.~(\ref{eom}) is one of the central results of this work and represents a diagrammatic generalization of carrier dynamics to dissipative multiband systems. As can be seen, in addition to dressing the known terms related to the Berry curvature and Levi-Civita connection, which are interpreted as the (multiband) momentum-space magnetic field and geodesic contributions, respectively, several additional contributions are identified. The remainder of this section is devoted to a discussion of these terms and their physical significance, particularly within the context of the gravitational interpretation.

The matrix $m^{ab} = \hbar/ \veps^{ab}$ (with
$\veps^{ab} \equiv \veps^a - \veps^b$) couples to the Christoffel symbol in the equation of motion and may thus be regarded as the local multiband generalization of the effective mass term in Ref.~\cite{smith2022momentum}. Physically, this implies that the effective gravitational force the electron in band $a$ experiences from the Levi-Civita connection between bands $b$ and $a$ is inversely proportional to the local energy difference between the two bands. Interestingly, this force can be attractive or repulsive depending on the sign of $\veps^{ab}$. And since the effective mass is a local quantity, its variation also contributes to the equation of motion, which explains the 
$\pd_{\mu} m^{ab}$ term in the equation of motion.

It is worth elaborating further on the appearance of dressed quantum geometric quantities in Eq.~(\ref{eom}) as opposed to bare ones. With the exception of the Levi-Civita connection, these are defined as
$\tilde{\mathcal{O}}^{ab}
\equiv
\lambda^{ab} \mathcal{O}^{ab}$,
where 
$\lambda^{ab}= [1+ (\eta^{ab})^2]^{-1}$, 
and $\eta^{ab}=\gamma/\veps^{ab}$ is a dimensionless measure of the dissipation strength, with $\gamma$ the self-energy~\footnote{We also note that this choice of dressing has the favorable property of preserving the $a \leftrightarrow b$ symmetries of the quantum metric and Berry curvature in band space.}. We stress that dressing the quantum geometry is an essential inclusion for a gravitational interpretation to hold, as, in general relativity, the matter distribution and the (spacetime) curvature are related and influence each other~\cite{misner1973gravitation}.

In our case, the introduction of scatterers to the system has the effect of screening the quantum geometry through the interband scattering function $\lambda^{ab}$, which acts as a dissipation-induced multiband Weyl transformation on the various geometric terms in the equation of motion. Physically, this accounts for the spectral broadening of the electron propagator from self-energy corrections due to scattering, and results in an effective rescaling of the two-state QGT in the linear response function. And the remaining dissipative corrections that arise in the quadratic responses can then be naturally absorbed in the renormalization functions, implying that the local ratios of the momentum-space forces an electron in a given band experiences from the various geometric terms are also affected by scattering, reflecting the nontrivial modification of the dynamics by dissipation.

The additional quantum geometric objects that appear in Eq.~(\ref{eom}) are given by
\begin{equation}
\mathcal{M}_{\mu \nu \rho}^{ba}
=
\mathcal{Z}_{K}^{ab}
\text{Re} (\tilde{K}_{\mu \nu \rho}^{ba})
+
\mathcal{Z}_{C}^{ab}
\text{Im}
\left(
\tilde{C}_{\mu \nu \rho}^{ba}
-
\tilde{C}_{\nu \mu \rho}^{ba}
-
\tilde{C}_{\rho \nu \mu}^{ba}
\right),
\end{equation}
which contains the contribution from the rest of the dressed Hermitian connection--including the contorsion--and
\begin{equation}
\label{N_munurho}
\begin{split}
\mathcal{N}_{\mu \nu \rho}^{abc}
&=
\mathcal{Z}_1^{abc} \text{Re} (\tilde{S}_{\mu\nu\rho}^{abc})
+
\mathcal{Z}_2^{abc} \text{Im} (\tilde{S}_{\mu\nu\rho}^{abc})
+
\mathcal{Z}_3^{abc} \text{Re} (\tilde{S}_{\nu\mu\rho}^{abc})
\\
&+
\mathcal{Z}_4^{abc} \text{Im} (\tilde{S}_{\nu\mu\rho}^{abc})
+
\mathcal{Z}_5^{abc} \text{Re} (\tilde{A}_{\nu\mu\rho}^{abc})
+
\mathcal{Z}_6^{abc} \text{Im} (\tilde{A}_{\nu\mu\rho}^{abc}),
\end{split}
\end{equation}
which encapsulates the contributions from the dressed three-state QGT~\footnote{Note that we define the dressed three-state QGT in terms of the first two band indices as $\tilde{Q}_{\mu\nu\rho}^{abc} = Q_{\mu\nu\rho}^{abc}/1+(\eta^{ab})^2$}. Together, $\mathcal{M}$ and  $\mathcal{N}$ can be intuitively viewed as describing intrinsic momentum-space matter fields~\cite{misner1973gravitation, wald2010general} that help capture the quantum geometry of Bloch electrons, including the symplectic and non-Abelian contributions, at second order in the carrier momentum.

Finally, consider now the term proportional to $\tilde{g}_{\mu\nu}^{ba}$ in Eq.~(\ref{eom}), which includes a dissipative linear-response addition to the equation of motion
\begin{equation}
\label{eq_drag}
\dot{x}_{\mu}^a
\supset
-2 \dot{k}^{\nu} \sum_b \eta^{ab} 
\tilde{g}_{\mu\nu}^{ba},
\end{equation}
that is also not captured in the semiclassical or density-matrix approaches. Applying the dual-space transformation $\mathbf{x} \leftrightarrow \mathbf{k}$ to this term reveals that, if the geodesic term is to be regarded as the momentum-space gravitational force, then the contribution from Eq.~(\ref{eq_drag}) can be thought of as arising from a dual drag force in momentum space induced by the scattering. In the next section, we discuss this more within the context of the EFE.

\textit{Dissipative Einstein field equations.}$-$We now discuss the effect of dissipation on the multiband momentum-space EFE and how it sources the field equations within the present formalism. In the absence of dissipation, the quantum metric is Fubini-Study, which is the canonical Riemannian metric of complex projective Hilbert space. This is an Einstein metric, with the Ricci tensor being proportional to the metric. Therefore, the bare metric satisfies the sourceless EFE, corresponding to the vacuum solution of the gravitational theory. Consider next the dissipative case. For convenience, we initially drop band indices from the notation, such that the indices $ab$ are implied for the various quantities that appear (i.e.,
$ g_{\mu\nu} \rightarrow g_{\mu\nu}^{ab}, R_{\mu\nu} \rightarrow R_{\mu\nu}^{ab}$, etc.). The inverse of the dressed metric is identified as
$\tilde{g}^{\mu\nu}= \lambda^{-1} g^{\mu\nu}$, such that
$\tilde{g}^{\mu \rho} \tilde{g}_{\rho \nu}
=
g^{\mu \rho} g_{\rho \nu}
=
\delta^{\mu}_{\nu}$, i.e., the dressed (bare) metric raises and lowers indices in the presence (absence) of dissipation. And the screened Levi-Civita connection is defined in the usual sense~\cite{wald2010general},
$\tilde{\nabla} \tilde{g} = 0$, which implies that the dressed Christoffel symbol is that of the dressed metric
\begin{equation}
\tilde{\Gamma}_{\mu\nu\rho}
=
\lambda \Gamma_{\mu\nu\rho}
+
\frac{1}{2} \left(
g_{\mu\nu}\nabla_{\rho} \lambda
+
g_{\mu\rho}\nabla_{\nu} \lambda
-
g_{\nu\rho}\nabla_{\mu} \lambda
\right).
\end{equation}
This can then be used to derive the dressed Riemann tensor
$\tensor{\tilde{R}}{^{\mu}_\nu_\rho_\sigma}$
and its contractions, namely the Ricci tensor
$\tilde{R}_{\mu\nu}
=
\tensor{\tilde{R}}{^{\rho}_\mu_\rho_\nu}$ and Ricci scalar
$\tilde{R}= \tensor{\tilde{R}}{^{\mu}_\mu}$.
Combining terms, we arrive at the dissipative EFE
\begin{equation}
\label{EFE}
\tilde{R}_{\mu\nu}
-\frac{1}{2} \tilde{R} \tilde{g}_{\mu\nu}
+
\Lambda \tilde{g}_{\mu\nu}
=
T_{\mu\nu},
\end{equation}
where $\Lambda$ and $T_{\mu\nu}$, which emerge entirely as a result of dissipative scattering, are identified as the local momentum-space cosmological constant and stress tensor, respectively, and are given, up to $O(\eta^2)$, by
\begin{subequations}
\begin{align}
\Lambda
&=
\nabla^2 \eta^2,
\\
T_{\mu\nu}
&=
\frac{n-2}{2}
\left( \nabla_{\mu} \nabla_{\nu} - g_{\mu\nu} \nabla^2 \right)\eta^2,
\end{align}
\end{subequations}
with $n$ the dimension of the system. This assignment is rather arbitrary as far as the EFE is concerned, and $\Lambda$ may well be absorbed into the stress tensor~\cite{misner1973gravitation}. What matters here instead is that the dressed Einstein tensor
$\tilde{G}_{\mu\nu}
=
\tilde{R}_{\mu\nu}
-
\tilde{R} \tilde{g}_{\mu\nu}/2$ does not vanish, which physically implies that dissipation is providing a source term for the momentum-space EFE.

To gain further insight into this result, it is helpful to recall that the modification of the EFE emerged from the screening of the metric. Following the semiclassical results, one could ask whether there is an entropic explanation for this. To answer this in a consistent manner within the present response-theory formalism, we note that, to second order in the electric field, the rate of local entropy production by microscopic scattering processes is determined by the symmetric part of the linear conductivity tensor~\cite{ziman2001electrons}
\begin{equation}
\dot{S}
=
\beta \mathbf{j} \cdot \mathbf{E}
=
\beta \sigma_{(\mu\nu)} E^{\mu} E^{\nu},
\end{equation}
which is related to Joule heating and has a quantum geometric contribution, $\dot{S}_{\text{g}}$, identified as giving rise to the momentum-space drag force, Eq.~(\ref{eq_drag}). This corresponds to the local conductivity contribution
$
- 2 e^2 / \hbar \sum_{ab}
f^a \eta^{ab} \tilde{g}_{\mu\nu}^{ab}
$
upon restoring band indices. Introducing the band resolution of the entropy rate as
$\dot{S} = \sum_{ab} \dot{S}^{ab}$, to $O(\eta^2)$ we find
\begin{equation}
\dot{S}_{\text{g}}^{ab}
=
- 2 \hbar \beta \dot{k}^{\mu} \dot{k}^{\nu} f^a g_{\mu\nu}^{ab}\eta^{ab}.
\end{equation}
This demonstrates that the source term in the multiband EFE  can be reexpressed as a function of the local entropy rate associated with dissipative interband processes, thereby providing a quantum-response generalization of the semiclassical arguments to dissipative multiband systems. Furthermore, in this sense, one can think of momentum-space gravity as essentially being a manifestation of local entropy changes in momentum-space, which is reminiscent of similar insights on gravity in spacetime. In this light, one could also think of the quantum geometric drag force related to Eq.~(\ref{eq_drag}) as a counterpart of the entropic force discussed in Ref.~\cite{verlinde2011origin}, with the difference that the latter--inspired by holographic arguments--is typically not associated with a fundamental field. In our case, this role is played by the quantum metric itself.

\textit{Conclusion and Outlook.}$-$In this work, we have presented an extension of semiclassical momentum-space gravity to dissipative multiband systems by taking quantum response theory as the starting point. Within a diagrammatic approach that includes a phenomenological dissipation parameter, we have studied the simultaneous and interconnected dressing of the quantum geometry, carrier dynamics and momentum-space EFE, resulting in a dressed theory of momentum-space gravity. On the technical side, this work introduces and provides two equivalent definitions of the three-state QGT as the simplest quantum geometric quantity beyond the two-state picture, which paves the way for future studies of higher-state quantum geometry. In addition, we have clarified the role of the quantum geometric contorsion tensor and its relation to the full Berry covariant derivative.

A general viewpoint on quantum response theory that has become more prominent recently is to view it as a probe of the rich Riemannian geometry of quantum state manifolds. The results presented here suggest that this may be harnessed to study a variety of theories of gravitation in quantum materials via optical, magnetic or thermal means once one identifies the corresponding gravitational effect in the multiband system. In addition, this approach shines an arguably more positive light on the role of dissipation, which is ubiquitous in any realistic system, but often regarded as an obstacle in probing intrinsic materials properties. Its interpretation as a generator of gravity in momentum space suggests a path to explore the connections between thermodynamics and gravity--which have been and continue to be of great interest in fundamental physics~\cite{bekenstein1973black, hawking1975particle, ruppeiner1979thermodynamics, jacobson1995thermodynamics, padmanabhan2010thermodynamical, verlinde2011origin, carroll2016what, bianconi2025gravity}--via responses in materials systems.

The analysis of the present work utilized standard Kubo formulas for closed quantum systems within the relaxation-time approximation, which suggests several generalizations for future works. One possibly interesting direction to explore from the gravitational point of view is the significance of special disorder scattering (i.e., side jump and skew scattering)--or nonscalar scatterings in the quantum geometry and carrier dynamics~\cite{mehraeen2024quantum, huang2025scaling, liu2024effect, gong2024nonlinear}. The former is particularly significant, given that special scattering processes in the nonlinear response regime often have relatively complicated and arguably unintuitive classifications. Therefore, finding their gravitational correspondences may be helpful in revealing a more intuitive picture of their nature. Furthermore, by considering system-bath correlations, one could generalize the present analysis to explore the interplay between quantum geometry and thermodynamics via momentum-space gravity in open quantum systems~\cite{konopik2019quantum}.
 
Finally, we anticipate that the results presented here may also be of general interest to studies in quantum science, information and technology. A notable direction in this regard is in the active field of quantum information geometry, given its common underlying geometric framework~\cite{facchi2010classical}. In addition, given recent developments in simulating dissipative nonlinear responses via generalizations of quantum phase estimation frameworks~\cite{loaiza2024nonlinear}, it would be quite interesting to consider the possibility of leveraging this approach to simulate theories of gravity via nonlinear responses on quantum computers.

\textit{Acknowledgments.}$-$We thank Ran Cheng, Michael Hinczewski, Kurt Hinterbichler, Wojciech Jankowski, Harsh Mathur, Robert-Jan Slager, Hantao Zhang and Shulei Zhang for helpful discussions.

\bibliography{qmsg}

\begin{thebibliography}{78}%
\makeatletter
\providecommand \@ifxundefined [1]{%
 \@ifx{#1\undefined}
}%
\providecommand \@ifnum [1]{%
 \ifnum #1\expandafter \@firstoftwo
 \else \expandafter \@secondoftwo
 \fi
}%
\providecommand \@ifx [1]{%
 \ifx #1\expandafter \@firstoftwo
 \else \expandafter \@secondoftwo
 \fi
}%
\providecommand \natexlab [1]{#1}%
\providecommand \enquote  [1]{``#1''}%
\providecommand \bibnamefont  [1]{#1}%
\providecommand \bibfnamefont [1]{#1}%
\providecommand \citenamefont [1]{#1}%
\providecommand \href@noop [0]{\@secondoftwo}%
\providecommand \href [0]{\begingroup \@sanitize@url \@href}%
\providecommand \@href[1]{\@@startlink{#1}\@@href}%
\providecommand \@@href[1]{\endgroup#1\@@endlink}%
\providecommand \@sanitize@url [0]{\catcode `\\12\catcode `\$12\catcode
  `\&12\catcode `\#12\catcode `\^12\catcode `\_12\catcode `\%12\relax}%
\providecommand \@@startlink[1]{}%
\providecommand \@@endlink[0]{}%
\providecommand \url  [0]{\begingroup\@sanitize@url \@url }%
\providecommand \@url [1]{\endgroup\@href {#1}{\urlprefix }}%
\providecommand \urlprefix  [0]{URL }%
\providecommand \Eprint [0]{\href }%
\providecommand \doibase [0]{https://doi.org/}%
\providecommand \selectlanguage [0]{\@gobble}%
\providecommand \bibinfo  [0]{\@secondoftwo}%
\providecommand \bibfield  [0]{\@secondoftwo}%
\providecommand \translation [1]{[#1]}%
\providecommand \BibitemOpen [0]{}%
\providecommand \bibitemStop [0]{}%
\providecommand \bibitemNoStop [0]{.\EOS\space}%
\providecommand \EOS [0]{\spacefactor3000\relax}%
\providecommand \BibitemShut  [1]{\csname bibitem#1\endcsname}%
\let\auto@bib@innerbib\@empty
\bibitem [{\citenamefont {Nakahara}(2018)}]{nakahara2018geometry}%
  \BibitemOpen
  \bibfield  {author} {\bibinfo {author} {\bibfnamefont {M.}~\bibnamefont
  {Nakahara}},\ }\href@noop {} {\emph {\bibinfo {title} {Geometry, topology and
  physics}}}\ (\bibinfo  {publisher} {CRC press},\ \bibinfo {year}
  {2018})\BibitemShut {NoStop}%
\bibitem [{\citenamefont {Ashtekar}\ and\ \citenamefont
  {Schilling}(1999)}]{ashtekar1999geometrical}%
  \BibitemOpen
  \bibfield  {author} {\bibinfo {author} {\bibfnamefont {A.}~\bibnamefont
  {Ashtekar}}\ and\ \bibinfo {author} {\bibfnamefont {T.~A.}\ \bibnamefont
  {Schilling}},\ }\bibfield  {title} {\bibinfo {title} {Geometrical formulation
  of quantum mechanics},\ }in\ \href@noop {} {\emph {\bibinfo {booktitle} {On
  Einstein’s Path: Essays in Honor of Engelbert Schucking}}}\ (\bibinfo
  {publisher} {Springer},\ \bibinfo {year} {1999})\ pp.\ \bibinfo {pages}
  {23--65}\BibitemShut {NoStop}%
\bibitem [{\citenamefont {Xiao}\ \emph {et~al.}(2010)\citenamefont {Xiao},
  \citenamefont {Chang},\ and\ \citenamefont {Niu}}]{xiao2010berry}%
  \BibitemOpen
  \bibfield  {author} {\bibinfo {author} {\bibfnamefont {D.}~\bibnamefont
  {Xiao}}, \bibinfo {author} {\bibfnamefont {M.-C.}\ \bibnamefont {Chang}},\
  and\ \bibinfo {author} {\bibfnamefont {Q.}~\bibnamefont {Niu}},\ }\bibfield
  {title} {\bibinfo {title} {{Berry phase effects on electronic properties}},\
  }\href {https://link.aps.org/doi/10.1103/RevModPhys.82.1959} {\bibfield
  {journal} {\bibinfo  {journal} {Rev. Mod. Phys.}\ }\textbf {\bibinfo {volume}
  {82}},\ \bibinfo {pages} {1959} (\bibinfo {year} {2010})}\BibitemShut
  {NoStop}%
\bibitem [{\citenamefont {T\"orm\"a}(2023)}]{torma2023essay}%
  \BibitemOpen
  \bibfield  {author} {\bibinfo {author} {\bibfnamefont {P.}~\bibnamefont
  {T\"orm\"a}},\ }\bibfield  {title} {\bibinfo {title} {{Essay: Where Can
  Quantum Geometry Lead Us?}},\ }\href
  {https://link.aps.org/doi/10.1103/PhysRevLett.131.240001} {\bibfield
  {journal} {\bibinfo  {journal} {Phys. Rev. Lett.}\ }\textbf {\bibinfo
  {volume} {131}},\ \bibinfo {pages} {240001} (\bibinfo {year}
  {2023})}\BibitemShut {NoStop}%
\bibitem [{\citenamefont {Liu}\ \emph {et~al.}(2024{\natexlab{a}})\citenamefont
  {Liu}, \citenamefont {Qiang}, \citenamefont {Lu},\ and\ \citenamefont
  {Xie}}]{liu2024quantum}%
  \BibitemOpen
  \bibfield  {author} {\bibinfo {author} {\bibfnamefont {T.}~\bibnamefont
  {Liu}}, \bibinfo {author} {\bibfnamefont {X.-B.}\ \bibnamefont {Qiang}},
  \bibinfo {author} {\bibfnamefont {H.-Z.}\ \bibnamefont {Lu}},\ and\ \bibinfo
  {author} {\bibfnamefont {X.}~\bibnamefont {Xie}},\ }\bibfield  {title}
  {\bibinfo {title} {{Quantum geometry in condensed matter}},\ }\href
  {https://doi.org/10.1093/nsr/nwae334} {\bibfield  {journal} {\bibinfo
  {journal} {Natl. Sci. Rev.}\ ,\ \bibinfo {pages} {nwae334}} (\bibinfo {year}
  {2024}{\natexlab{a}})}\BibitemShut {NoStop}%
\bibitem [{\citenamefont {Yu}\ \emph {et~al.}(2024)\citenamefont {Yu},
  \citenamefont {Bernevig}, \citenamefont {Queiroz}, \citenamefont {Rossi},
  \citenamefont {T{\"o}rm{\"a}},\ and\ \citenamefont {Yang}}]{yu2024quantum}%
  \BibitemOpen
  \bibfield  {author} {\bibinfo {author} {\bibfnamefont {J.}~\bibnamefont
  {Yu}}, \bibinfo {author} {\bibfnamefont {B.~A.}\ \bibnamefont {Bernevig}},
  \bibinfo {author} {\bibfnamefont {R.}~\bibnamefont {Queiroz}}, \bibinfo
  {author} {\bibfnamefont {E.}~\bibnamefont {Rossi}}, \bibinfo {author}
  {\bibfnamefont {P.}~\bibnamefont {T{\"o}rm{\"a}}},\ and\ \bibinfo {author}
  {\bibfnamefont {B.-J.}\ \bibnamefont {Yang}},\ }\bibfield  {title} {\bibinfo
  {title} {{Quantum Geometry in Quantum Materials}},\ }\href
  {https://arxiv.org/abs/2501.00098} {\bibfield  {journal} {\bibinfo  {journal}
  {arXiv:2501.00098}\ } (\bibinfo {year} {2024})}\BibitemShut {NoStop}%
\bibitem [{\citenamefont {Chen}(2024)}]{chen2024quantum}%
  \BibitemOpen
  \bibfield  {author} {\bibinfo {author} {\bibfnamefont {W.}~\bibnamefont
  {Chen}},\ }\bibfield  {title} {\bibinfo {title} {Quantum geometrical
  properties of topological materials},\ }\href
  {http://dx.doi.org/10.1088/1361-648X/ad8619} {\bibfield  {journal} {\bibinfo
  {journal} {J. Phys.: Condens. Matter}\ }\textbf {\bibinfo {volume} {37}},\
  \bibinfo {pages} {025605} (\bibinfo {year} {2024})}\BibitemShut {NoStop}%
\bibitem [{\citenamefont {Shim}\ \emph {et~al.}(2025)\citenamefont {Shim},
  \citenamefont {Mehraeen}, \citenamefont {Sklenar}, \citenamefont {Zhang},
  \citenamefont {Hoffmann},\ and\ \citenamefont {Mason}}]{shim2025spin}%
  \BibitemOpen
  \bibfield  {author} {\bibinfo {author} {\bibfnamefont {S.}~\bibnamefont
  {Shim}}, \bibinfo {author} {\bibfnamefont {M.}~\bibnamefont {Mehraeen}},
  \bibinfo {author} {\bibfnamefont {J.}~\bibnamefont {Sklenar}}, \bibinfo
  {author} {\bibfnamefont {S.~S.-L.}\ \bibnamefont {Zhang}}, \bibinfo {author}
  {\bibfnamefont {A.}~\bibnamefont {Hoffmann}},\ and\ \bibinfo {author}
  {\bibfnamefont {N.}~\bibnamefont {Mason}},\ }\bibfield  {title} {\bibinfo
  {title} {Spin-polarized antiferromagnetic metals},\ }\href
  {https://doi.org/10.1146/annurev-conmatphys-042924-123620} {\bibfield
  {journal} {\bibinfo  {journal} {Annu. Rev. Condens. Matter Phys.}\ }\textbf
  {\bibinfo {volume} {16}} (\bibinfo {year} {2025})}\BibitemShut {NoStop}%
\bibitem [{\citenamefont {Jiang}\ \emph {et~al.}(2025)\citenamefont {Jiang},
  \citenamefont {Holder},\ and\ \citenamefont {Yan}}]{jiang2025revealing}%
  \BibitemOpen
  \bibfield  {author} {\bibinfo {author} {\bibfnamefont {Y.}~\bibnamefont
  {Jiang}}, \bibinfo {author} {\bibfnamefont {T.}~\bibnamefont {Holder}},\ and\
  \bibinfo {author} {\bibfnamefont {B.}~\bibnamefont {Yan}},\ }\bibfield
  {title} {\bibinfo {title} {{Revealing quantum geometry in nonlinear quantum
  materials}},\ }\href {https://doi.org/10.1088/1361-6633/ade454} {\bibfield
  {journal} {\bibinfo  {journal} {Rep. Prog. Phys.}\ }\textbf {\bibinfo
  {volume} {88}} (\bibinfo {year} {2025})}\BibitemShut {NoStop}%
\bibitem [{\citenamefont {Verma}\ \emph {et~al.}(2025)\citenamefont {Verma},
  \citenamefont {Moll}, \citenamefont {Holder},\ and\ \citenamefont
  {Queiroz}}]{verma2025quantum}%
  \BibitemOpen
  \bibfield  {author} {\bibinfo {author} {\bibfnamefont {N.}~\bibnamefont
  {Verma}}, \bibinfo {author} {\bibfnamefont {P.~J.}\ \bibnamefont {Moll}},
  \bibinfo {author} {\bibfnamefont {T.}~\bibnamefont {Holder}},\ and\ \bibinfo
  {author} {\bibfnamefont {R.}~\bibnamefont {Queiroz}},\ }\bibfield  {title}
  {\bibinfo {title} {{Quantum Geometry: Revisiting electronic scales in quantum
  matter}},\ }\href {https://arxiv.org/abs/2504.07173} {\bibfield  {journal}
  {\bibinfo  {journal} {arXiv:2504.07173}\ } (\bibinfo {year}
  {2025})}\BibitemShut {NoStop}%
\bibitem [{\citenamefont {Provost}\ and\ \citenamefont
  {Vallee}(1980)}]{provost1980riemannian}%
  \BibitemOpen
  \bibfield  {author} {\bibinfo {author} {\bibfnamefont {J.}~\bibnamefont
  {Provost}}\ and\ \bibinfo {author} {\bibfnamefont {G.}~\bibnamefont
  {Vallee}},\ }\bibfield  {title} {\bibinfo {title} {{Riemannian structure on
  manifolds of quantum states}},\ }\href {https://doi.org/10.1007/BF02193559}
  {\bibfield  {journal} {\bibinfo  {journal} {Commun. Math. Phys.}\ }\textbf
  {\bibinfo {volume} {76}},\ \bibinfo {pages} {289} (\bibinfo {year}
  {1980})}\BibitemShut {NoStop}%
\bibitem [{\citenamefont {Neupert}\ \emph {et~al.}(2013)\citenamefont
  {Neupert}, \citenamefont {Chamon},\ and\ \citenamefont
  {Mudry}}]{neupert2013measuring}%
  \BibitemOpen
  \bibfield  {author} {\bibinfo {author} {\bibfnamefont {T.}~\bibnamefont
  {Neupert}}, \bibinfo {author} {\bibfnamefont {C.}~\bibnamefont {Chamon}},\
  and\ \bibinfo {author} {\bibfnamefont {C.}~\bibnamefont {Mudry}},\ }\bibfield
   {title} {\bibinfo {title} {{Measuring the quantum geometry of Bloch bands
  with current noise}},\ }\href
  {https://link.aps.org/doi/10.1103/PhysRevB.87.245103} {\bibfield  {journal}
  {\bibinfo  {journal} {Phys. Rev. B}\ }\textbf {\bibinfo {volume} {87}},\
  \bibinfo {pages} {245103} (\bibinfo {year} {2013})}\BibitemShut {NoStop}%
\bibitem [{\citenamefont {Claassen}\ \emph {et~al.}(2015)\citenamefont
  {Claassen}, \citenamefont {Lee}, \citenamefont {Thomale}, \citenamefont
  {Qi},\ and\ \citenamefont {Devereaux}}]{claassen2015position}%
  \BibitemOpen
  \bibfield  {author} {\bibinfo {author} {\bibfnamefont {M.}~\bibnamefont
  {Claassen}}, \bibinfo {author} {\bibfnamefont {C.~H.}\ \bibnamefont {Lee}},
  \bibinfo {author} {\bibfnamefont {R.}~\bibnamefont {Thomale}}, \bibinfo
  {author} {\bibfnamefont {X.-L.}\ \bibnamefont {Qi}},\ and\ \bibinfo {author}
  {\bibfnamefont {T.~P.}\ \bibnamefont {Devereaux}},\ }\bibfield  {title}
  {\bibinfo {title} {{Position-Momentum Duality and Fractional Quantum Hall
  Effect in Chern Insulators}},\ }\href
  {https://link.aps.org/doi/10.1103/PhysRevLett.114.236802} {\bibfield
  {journal} {\bibinfo  {journal} {Phys. Rev. Lett.}\ }\textbf {\bibinfo
  {volume} {114}},\ \bibinfo {pages} {236802} (\bibinfo {year}
  {2015})}\BibitemShut {NoStop}%
\bibitem [{\citenamefont {Ahn}\ \emph {et~al.}(2020)\citenamefont {Ahn},
  \citenamefont {Guo},\ and\ \citenamefont {Nagaosa}}]{ahn2020low}%
  \BibitemOpen
  \bibfield  {author} {\bibinfo {author} {\bibfnamefont {J.}~\bibnamefont
  {Ahn}}, \bibinfo {author} {\bibfnamefont {G.-Y.}\ \bibnamefont {Guo}},\ and\
  \bibinfo {author} {\bibfnamefont {N.}~\bibnamefont {Nagaosa}},\ }\bibfield
  {title} {\bibinfo {title} {{Low-Frequency Divergence and Quantum Geometry of
  the Bulk Photovoltaic Effect in Topological Semimetals}},\ }\href
  {https://link.aps.org/doi/10.1103/PhysRevX.10.041041} {\bibfield  {journal}
  {\bibinfo  {journal} {Phys. Rev. X}\ }\textbf {\bibinfo {volume} {10}},\
  \bibinfo {pages} {041041} (\bibinfo {year} {2020})}\BibitemShut {NoStop}%
\bibitem [{\citenamefont {Watanabe}\ and\ \citenamefont
  {Yanase}(2021)}]{watanabe2021chiral}%
  \BibitemOpen
  \bibfield  {author} {\bibinfo {author} {\bibfnamefont {H.}~\bibnamefont
  {Watanabe}}\ and\ \bibinfo {author} {\bibfnamefont {Y.}~\bibnamefont
  {Yanase}},\ }\bibfield  {title} {\bibinfo {title} {{Chiral Photocurrent in
  Parity-Violating Magnet and Enhanced Response in Topological
  Antiferromagnet}},\ }\href
  {https://link.aps.org/doi/10.1103/PhysRevX.11.011001} {\bibfield  {journal}
  {\bibinfo  {journal} {Phys. Rev. X}\ }\textbf {\bibinfo {volume} {11}},\
  \bibinfo {pages} {011001} (\bibinfo {year} {2021})}\BibitemShut {NoStop}%
\bibitem [{\citenamefont {Wang}\ \emph {et~al.}(2021)\citenamefont {Wang},
  \citenamefont {Gao},\ and\ \citenamefont {Xiao}}]{wang2021intrinsic}%
  \BibitemOpen
  \bibfield  {author} {\bibinfo {author} {\bibfnamefont {C.}~\bibnamefont
  {Wang}}, \bibinfo {author} {\bibfnamefont {Y.}~\bibnamefont {Gao}},\ and\
  \bibinfo {author} {\bibfnamefont {D.}~\bibnamefont {Xiao}},\ }\bibfield
  {title} {\bibinfo {title} {{Intrinsic Nonlinear \text{H}all Effect in
  Antiferromagnetic Tetragonal CuMnAs}},\ }\href
  {https://link.aps.org/doi/10.1103/PhysRevLett.127.277201} {\bibfield
  {journal} {\bibinfo  {journal} {Phys. Rev. Lett.}\ }\textbf {\bibinfo
  {volume} {127}},\ \bibinfo {pages} {277201} (\bibinfo {year}
  {2021})}\BibitemShut {NoStop}%
\bibitem [{\citenamefont {Liu}\ \emph {et~al.}(2021)\citenamefont {Liu},
  \citenamefont {Zhao}, \citenamefont {Huang}, \citenamefont {Wu},
  \citenamefont {Sheng}, \citenamefont {Xiao},\ and\ \citenamefont
  {Yang}}]{liu2021intrinsic}%
  \BibitemOpen
  \bibfield  {author} {\bibinfo {author} {\bibfnamefont {H.}~\bibnamefont
  {Liu}}, \bibinfo {author} {\bibfnamefont {J.}~\bibnamefont {Zhao}}, \bibinfo
  {author} {\bibfnamefont {Y.-X.}\ \bibnamefont {Huang}}, \bibinfo {author}
  {\bibfnamefont {W.}~\bibnamefont {Wu}}, \bibinfo {author} {\bibfnamefont
  {X.-L.}\ \bibnamefont {Sheng}}, \bibinfo {author} {\bibfnamefont
  {C.}~\bibnamefont {Xiao}},\ and\ \bibinfo {author} {\bibfnamefont {S.~A.}\
  \bibnamefont {Yang}},\ }\bibfield  {title} {\bibinfo {title} {{Intrinsic
  Second-Order Anomalous Hall Effect and Its Application in Compensated
  Antiferromagnets}},\ }\href
  {https://link.aps.org/doi/10.1103/PhysRevLett.127.277202} {\bibfield
  {journal} {\bibinfo  {journal} {Phys. Rev. Lett.}\ }\textbf {\bibinfo
  {volume} {127}},\ \bibinfo {pages} {277202} (\bibinfo {year}
  {2021})}\BibitemShut {NoStop}%
\bibitem [{\citenamefont {Bhalla}\ \emph {et~al.}(2022)\citenamefont {Bhalla},
  \citenamefont {Das}, \citenamefont {Culcer},\ and\ \citenamefont
  {Agarwal}}]{bhalla2022resonant}%
  \BibitemOpen
  \bibfield  {author} {\bibinfo {author} {\bibfnamefont {P.}~\bibnamefont
  {Bhalla}}, \bibinfo {author} {\bibfnamefont {K.}~\bibnamefont {Das}},
  \bibinfo {author} {\bibfnamefont {D.}~\bibnamefont {Culcer}},\ and\ \bibinfo
  {author} {\bibfnamefont {A.}~\bibnamefont {Agarwal}},\ }\bibfield  {title}
  {\bibinfo {title} {{Resonant Second-Harmonic Generation as a Probe of Quantum
  Geometry}},\ }\href {https://link.aps.org/doi/10.1103/PhysRevLett.129.227401}
  {\bibfield  {journal} {\bibinfo  {journal} {Phys. Rev. Lett.}\ }\textbf
  {\bibinfo {volume} {129}},\ \bibinfo {pages} {227401} (\bibinfo {year}
  {2022})}\BibitemShut {NoStop}%
\bibitem [{\citenamefont {Gao}\ \emph {et~al.}(2023)\citenamefont {Gao},
  \citenamefont {Liu}, \citenamefont {Qiu}, \citenamefont {Ghosh},
  \citenamefont {V.~Trevisan}, \citenamefont {Onishi}, \citenamefont {Hu},
  \citenamefont {Qian}, \citenamefont {Tien}, \citenamefont {Chen} \emph
  {et~al.}}]{gao2023quantum}%
  \BibitemOpen
  \bibfield  {author} {\bibinfo {author} {\bibfnamefont {A.}~\bibnamefont
  {Gao}}, \bibinfo {author} {\bibfnamefont {Y.-F.}\ \bibnamefont {Liu}},
  \bibinfo {author} {\bibfnamefont {J.-X.}\ \bibnamefont {Qiu}}, \bibinfo
  {author} {\bibfnamefont {B.}~\bibnamefont {Ghosh}}, \bibinfo {author}
  {\bibfnamefont {T.}~\bibnamefont {V.~Trevisan}}, \bibinfo {author}
  {\bibfnamefont {Y.}~\bibnamefont {Onishi}}, \bibinfo {author} {\bibfnamefont
  {C.}~\bibnamefont {Hu}}, \bibinfo {author} {\bibfnamefont {T.}~\bibnamefont
  {Qian}}, \bibinfo {author} {\bibfnamefont {H.-J.}\ \bibnamefont {Tien}},
  \bibinfo {author} {\bibfnamefont {S.-W.}\ \bibnamefont {Chen}}, \emph
  {et~al.},\ }\bibfield  {title} {\bibinfo {title} {{Quantum metric nonlinear
  Hall effect in a topological antiferromagnetic heterostructure}},\ }\href
  {https://doi.org/10.1126/science.adf1506} {\bibfield  {journal} {\bibinfo
  {journal} {Science}\ ,\ \bibinfo {pages} {eadf1506}} (\bibinfo {year}
  {2023})}\BibitemShut {NoStop}%
\bibitem [{\citenamefont {Wang}\ \emph {et~al.}(2023)\citenamefont {Wang},
  \citenamefont {Kaplan}, \citenamefont {Zhang}, \citenamefont {Holder},
  \citenamefont {Cao}, \citenamefont {Wang}, \citenamefont {Zhou},
  \citenamefont {Zhou}, \citenamefont {Jiang}, \citenamefont {Zhang} \emph
  {et~al.}}]{wang2023quantum}%
  \BibitemOpen
  \bibfield  {author} {\bibinfo {author} {\bibfnamefont {N.}~\bibnamefont
  {Wang}}, \bibinfo {author} {\bibfnamefont {D.}~\bibnamefont {Kaplan}},
  \bibinfo {author} {\bibfnamefont {Z.}~\bibnamefont {Zhang}}, \bibinfo
  {author} {\bibfnamefont {T.}~\bibnamefont {Holder}}, \bibinfo {author}
  {\bibfnamefont {N.}~\bibnamefont {Cao}}, \bibinfo {author} {\bibfnamefont
  {A.}~\bibnamefont {Wang}}, \bibinfo {author} {\bibfnamefont {X.}~\bibnamefont
  {Zhou}}, \bibinfo {author} {\bibfnamefont {F.}~\bibnamefont {Zhou}}, \bibinfo
  {author} {\bibfnamefont {Z.}~\bibnamefont {Jiang}}, \bibinfo {author}
  {\bibfnamefont {C.}~\bibnamefont {Zhang}}, \emph {et~al.},\ }\bibfield
  {title} {\bibinfo {title} {{Quantum-metric-induced nonlinear transport in a
  topological antiferromagnet}},\ }\href
  {https://doi.org/10.1038/s41586-023-06363-3} {\bibfield  {journal} {\bibinfo
  {journal} {Nature}\ }\textbf {\bibinfo {volume} {621}},\ \bibinfo {pages}
  {487} (\bibinfo {year} {2023})}\BibitemShut {NoStop}%
\bibitem [{\citenamefont {Het\'enyi}\ and\ \citenamefont
  {L\'evay}(2023)}]{hetenyi2023fluctuations}%
  \BibitemOpen
  \bibfield  {author} {\bibinfo {author} {\bibfnamefont {B.}~\bibnamefont
  {Het\'enyi}}\ and\ \bibinfo {author} {\bibfnamefont {P.}~\bibnamefont
  {L\'evay}},\ }\bibfield  {title} {\bibinfo {title} {{Fluctuations,
  uncertainty relations, and the geometry of quantum state manifolds}},\ }\href
  {https://link.aps.org/doi/10.1103/PhysRevA.108.032218} {\bibfield  {journal}
  {\bibinfo  {journal} {Phys. Rev. A}\ }\textbf {\bibinfo {volume} {108}},\
  \bibinfo {pages} {032218} (\bibinfo {year} {2023})}\BibitemShut {NoStop}%
\bibitem [{\citenamefont {Das}\ \emph {et~al.}(2023)\citenamefont {Das},
  \citenamefont {Lahiri}, \citenamefont {Atencia}, \citenamefont {Culcer},\
  and\ \citenamefont {Agarwal}}]{das2023intrinsic}%
  \BibitemOpen
  \bibfield  {author} {\bibinfo {author} {\bibfnamefont {K.}~\bibnamefont
  {Das}}, \bibinfo {author} {\bibfnamefont {S.}~\bibnamefont {Lahiri}},
  \bibinfo {author} {\bibfnamefont {R.~B.}\ \bibnamefont {Atencia}}, \bibinfo
  {author} {\bibfnamefont {D.}~\bibnamefont {Culcer}},\ and\ \bibinfo {author}
  {\bibfnamefont {A.}~\bibnamefont {Agarwal}},\ }\bibfield  {title} {\bibinfo
  {title} {{Intrinsic nonlinear conductivities induced by the quantum
  metric}},\ }\href {https://link.aps.org/doi/10.1103/PhysRevB.108.L201405}
  {\bibfield  {journal} {\bibinfo  {journal} {Phys. Rev. B}\ }\textbf {\bibinfo
  {volume} {108}},\ \bibinfo {pages} {L201405} (\bibinfo {year}
  {2023})}\BibitemShut {NoStop}%
\bibitem [{\citenamefont {Komissarov}\ \emph {et~al.}(2024)\citenamefont
  {Komissarov}, \citenamefont {Holder},\ and\ \citenamefont
  {Queiroz}}]{komissarov2024quantum}%
  \BibitemOpen
  \bibfield  {author} {\bibinfo {author} {\bibfnamefont {I.}~\bibnamefont
  {Komissarov}}, \bibinfo {author} {\bibfnamefont {T.}~\bibnamefont {Holder}},\
  and\ \bibinfo {author} {\bibfnamefont {R.}~\bibnamefont {Queiroz}},\
  }\bibfield  {title} {\bibinfo {title} {The quantum geometric origin of
  capacitance in insulators},\ }\href
  {https://doi.org/10.1038/s41467-024-48808-x} {\bibfield  {journal} {\bibinfo
  {journal} {Nat. Commun.}\ }\textbf {\bibinfo {volume} {15}},\ \bibinfo
  {pages} {4621} (\bibinfo {year} {2024})}\BibitemShut {NoStop}%
\bibitem [{\citenamefont {Onishi}\ and\ \citenamefont
  {Fu}(2024)}]{onishi2024fundamental}%
  \BibitemOpen
  \bibfield  {author} {\bibinfo {author} {\bibfnamefont {Y.}~\bibnamefont
  {Onishi}}\ and\ \bibinfo {author} {\bibfnamefont {L.}~\bibnamefont {Fu}},\
  }\bibfield  {title} {\bibinfo {title} {Fundamental bound on topological
  gap},\ }\href {https://link.aps.org/doi/10.1103/PhysRevX.14.011052}
  {\bibfield  {journal} {\bibinfo  {journal} {Phys. Rev. X}\ }\textbf {\bibinfo
  {volume} {14}},\ \bibinfo {pages} {011052} (\bibinfo {year}
  {2024})}\BibitemShut {NoStop}%
\bibitem [{\citenamefont {Fang}\ \emph {et~al.}(2024)\citenamefont {Fang},
  \citenamefont {Cano},\ and\ \citenamefont {Ghorashi}}]{fang2024quantum}%
  \BibitemOpen
  \bibfield  {author} {\bibinfo {author} {\bibfnamefont {Y.}~\bibnamefont
  {Fang}}, \bibinfo {author} {\bibfnamefont {J.}~\bibnamefont {Cano}},\ and\
  \bibinfo {author} {\bibfnamefont {S.~A.~A.}\ \bibnamefont {Ghorashi}},\
  }\bibfield  {title} {\bibinfo {title} {{Quantum Geometry Induced Nonlinear
  Transport in Altermagnets}},\ }\href
  {https://link.aps.org/doi/10.1103/PhysRevLett.133.106701} {\bibfield
  {journal} {\bibinfo  {journal} {Phys. Rev. Lett.}\ }\textbf {\bibinfo
  {volume} {133}},\ \bibinfo {pages} {106701} (\bibinfo {year}
  {2024})}\BibitemShut {NoStop}%
\bibitem [{\citenamefont {Kang}\ \emph {et~al.}(2024)\citenamefont {Kang},
  \citenamefont {Kim}, \citenamefont {Qian}, \citenamefont {Neves},
  \citenamefont {Ye}, \citenamefont {Jung}, \citenamefont {Puntel},
  \citenamefont {Mazzola}, \citenamefont {Fang}, \citenamefont {Jozwiak} \emph
  {et~al.}}]{kang2024measurements}%
  \BibitemOpen
  \bibfield  {author} {\bibinfo {author} {\bibfnamefont {M.}~\bibnamefont
  {Kang}}, \bibinfo {author} {\bibfnamefont {S.}~\bibnamefont {Kim}}, \bibinfo
  {author} {\bibfnamefont {Y.}~\bibnamefont {Qian}}, \bibinfo {author}
  {\bibfnamefont {P.~M.}\ \bibnamefont {Neves}}, \bibinfo {author}
  {\bibfnamefont {L.}~\bibnamefont {Ye}}, \bibinfo {author} {\bibfnamefont
  {J.}~\bibnamefont {Jung}}, \bibinfo {author} {\bibfnamefont {D.}~\bibnamefont
  {Puntel}}, \bibinfo {author} {\bibfnamefont {F.}~\bibnamefont {Mazzola}},
  \bibinfo {author} {\bibfnamefont {S.}~\bibnamefont {Fang}}, \bibinfo {author}
  {\bibfnamefont {C.}~\bibnamefont {Jozwiak}}, \emph {et~al.},\ }\bibfield
  {title} {\bibinfo {title} {Measurements of the quantum geometric tensor in
  solids},\ }\href {https://doi.org/10.1038/s41567-024-02678-8} {\bibfield
  {journal} {\bibinfo  {journal} {Nat. Phys.}\ ,\ \bibinfo {pages} {1}}
  (\bibinfo {year} {2024})}\BibitemShut {NoStop}%
\bibitem [{\citenamefont {Jankowski}\ \emph
  {et~al.}(2025{\natexlab{a}})\citenamefont {Jankowski}, \citenamefont
  {Morris}, \citenamefont {Bouhon}, \citenamefont {\"Unal},\ and\ \citenamefont
  {Slager}}]{jankowski2025optical}%
  \BibitemOpen
  \bibfield  {author} {\bibinfo {author} {\bibfnamefont {W.~J.}\ \bibnamefont
  {Jankowski}}, \bibinfo {author} {\bibfnamefont {A.~S.}\ \bibnamefont
  {Morris}}, \bibinfo {author} {\bibfnamefont {A.}~\bibnamefont {Bouhon}},
  \bibinfo {author} {\bibfnamefont {F.~N.}\ \bibnamefont {\"Unal}},\ and\
  \bibinfo {author} {\bibfnamefont {R.-J.}\ \bibnamefont {Slager}},\ }\bibfield
   {title} {\bibinfo {title} {{Optical manifestations and bounds of topological
  Euler class}},\ }\href
  {https://link.aps.org/doi/10.1103/PhysRevB.111.L081103} {\bibfield  {journal}
  {\bibinfo  {journal} {Phys. Rev. B}\ }\textbf {\bibinfo {volume} {111}},\
  \bibinfo {pages} {L081103} (\bibinfo {year}
  {2025}{\natexlab{a}})}\BibitemShut {NoStop}%
\bibitem [{\citenamefont {Karplus}\ and\ \citenamefont
  {Luttinger}(1954)}]{karplus1954hall}%
  \BibitemOpen
  \bibfield  {author} {\bibinfo {author} {\bibfnamefont {R.}~\bibnamefont
  {Karplus}}\ and\ \bibinfo {author} {\bibfnamefont {J.~M.}\ \bibnamefont
  {Luttinger}},\ }\bibfield  {title} {\bibinfo {title} {{Hall Effect in
  Ferromagnetics}},\ }\href {https://link.aps.org/doi/10.1103/PhysRev.95.1154}
  {\bibfield  {journal} {\bibinfo  {journal} {Phys. Rev.}\ }\textbf {\bibinfo
  {volume} {95}},\ \bibinfo {pages} {1154} (\bibinfo {year}
  {1954})}\BibitemShut {NoStop}%
\bibitem [{\citenamefont {Kohn}\ and\ \citenamefont
  {Luttinger}(1957)}]{kohn1957quantum}%
  \BibitemOpen
  \bibfield  {author} {\bibinfo {author} {\bibfnamefont {W.}~\bibnamefont
  {Kohn}}\ and\ \bibinfo {author} {\bibfnamefont {J.~M.}\ \bibnamefont
  {Luttinger}},\ }\bibfield  {title} {\bibinfo {title} {{Quantum Theory of
  Electrical Transport Phenomena}},\ }\href
  {https://link.aps.org/doi/10.1103/PhysRev.108.590} {\bibfield  {journal}
  {\bibinfo  {journal} {Phys. Rev.}\ }\textbf {\bibinfo {volume} {108}},\
  \bibinfo {pages} {590} (\bibinfo {year} {1957})}\BibitemShut {NoStop}%
\bibitem [{\citenamefont {Adams}\ and\ \citenamefont
  {Blount}(1959)}]{adams1959energy}%
  \BibitemOpen
  \bibfield  {author} {\bibinfo {author} {\bibfnamefont {E.}~\bibnamefont
  {Adams}}\ and\ \bibinfo {author} {\bibfnamefont {E.}~\bibnamefont {Blount}},\
  }\bibfield  {title} {\bibinfo {title} {{Energy bands in the presence of an
  external force field—II: Anomalous velocities}},\ }\href
  {https://doi.org/10.1016/0022-3697(59)90004-6} {\bibfield  {journal}
  {\bibinfo  {journal} {J. Phys. Chem. Solids}\ }\textbf {\bibinfo {volume}
  {10}},\ \bibinfo {pages} {286} (\bibinfo {year} {1959})}\BibitemShut
  {NoStop}%
\bibitem [{\citenamefont {Chang}\ and\ \citenamefont
  {Niu}(1995)}]{chang1995berry}%
  \BibitemOpen
  \bibfield  {author} {\bibinfo {author} {\bibfnamefont {M.-C.}\ \bibnamefont
  {Chang}}\ and\ \bibinfo {author} {\bibfnamefont {Q.}~\bibnamefont {Niu}},\
  }\bibfield  {title} {\bibinfo {title} {{Berry Phase, Hyperorbits, and the
  Hofstadter Spectrum}},\ }\href
  {https://link.aps.org/doi/10.1103/PhysRevLett.75.1348} {\bibfield  {journal}
  {\bibinfo  {journal} {Phys. Rev. Lett.}\ }\textbf {\bibinfo {volume} {75}},\
  \bibinfo {pages} {1348} (\bibinfo {year} {1995})}\BibitemShut {NoStop}%
\bibitem [{\citenamefont {Chang}\ and\ \citenamefont
  {Niu}(1996)}]{chang1996berry}%
  \BibitemOpen
  \bibfield  {author} {\bibinfo {author} {\bibfnamefont {M.-C.}\ \bibnamefont
  {Chang}}\ and\ \bibinfo {author} {\bibfnamefont {Q.}~\bibnamefont {Niu}},\
  }\bibfield  {title} {\bibinfo {title} {{Berry phase, hyperorbits, and the
  Hofstadter spectrum: Semiclassical dynamics in magnetic Bloch bands}},\
  }\href {https://link.aps.org/doi/10.1103/PhysRevB.53.7010} {\bibfield
  {journal} {\bibinfo  {journal} {Phys. Rev. B}\ }\textbf {\bibinfo {volume}
  {53}},\ \bibinfo {pages} {7010} (\bibinfo {year} {1996})}\BibitemShut
  {NoStop}%
\bibitem [{\citenamefont {Sundaram}\ and\ \citenamefont
  {Niu}(1999)}]{sundaram1999wave}%
  \BibitemOpen
  \bibfield  {author} {\bibinfo {author} {\bibfnamefont {G.}~\bibnamefont
  {Sundaram}}\ and\ \bibinfo {author} {\bibfnamefont {Q.}~\bibnamefont {Niu}},\
  }\bibfield  {title} {\bibinfo {title} {{Wave-packet dynamics in slowly
  perturbed crystals: Gradient corrections and Berry-phase effects}},\ }\href
  {https://link.aps.org/doi/10.1103/PhysRevB.59.14915} {\bibfield  {journal}
  {\bibinfo  {journal} {Phys. Rev. B}\ }\textbf {\bibinfo {volume} {59}},\
  \bibinfo {pages} {14915} (\bibinfo {year} {1999})}\BibitemShut {NoStop}%
\bibitem [{\citenamefont {Gao}\ \emph {et~al.}(2014)\citenamefont {Gao},
  \citenamefont {Yang},\ and\ \citenamefont {Niu}}]{gao2014field}%
  \BibitemOpen
  \bibfield  {author} {\bibinfo {author} {\bibfnamefont {Y.}~\bibnamefont
  {Gao}}, \bibinfo {author} {\bibfnamefont {S.~A.}\ \bibnamefont {Yang}},\ and\
  \bibinfo {author} {\bibfnamefont {Q.}~\bibnamefont {Niu}},\ }\bibfield
  {title} {\bibinfo {title} {{Field Induced Positional Shift of Bloch Electrons
  and Its Dynamical Implications}},\ }\href
  {https://link.aps.org/doi/10.1103/PhysRevLett.112.166601} {\bibfield
  {journal} {\bibinfo  {journal} {Phys. Rev. Lett.}\ }\textbf {\bibinfo
  {volume} {112}},\ \bibinfo {pages} {166601} (\bibinfo {year}
  {2014})}\BibitemShut {NoStop}%
\bibitem [{\citenamefont {Smith}\ \emph {et~al.}(2022)\citenamefont {Smith},
  \citenamefont {Pullasseri},\ and\ \citenamefont
  {Srivastava}}]{smith2022momentum}%
  \BibitemOpen
  \bibfield  {author} {\bibinfo {author} {\bibfnamefont {T.~B.}\ \bibnamefont
  {Smith}}, \bibinfo {author} {\bibfnamefont {L.}~\bibnamefont {Pullasseri}},\
  and\ \bibinfo {author} {\bibfnamefont {A.}~\bibnamefont {Srivastava}},\
  }\bibfield  {title} {\bibinfo {title} {{Momentum-space gravity from the
  quantum geometry and entropy of Bloch electrons}},\ }\href
  {https://link.aps.org/doi/10.1103/PhysRevResearch.4.013217} {\bibfield
  {journal} {\bibinfo  {journal} {Phys. Rev. Res.}\ }\textbf {\bibinfo {volume}
  {4}},\ \bibinfo {pages} {013217} (\bibinfo {year} {2022})}\BibitemShut
  {NoStop}%
\bibitem [{\citenamefont {Bekenstein}(1973)}]{bekenstein1973black}%
  \BibitemOpen
  \bibfield  {author} {\bibinfo {author} {\bibfnamefont {J.~D.}\ \bibnamefont
  {Bekenstein}},\ }\bibfield  {title} {\bibinfo {title} {{Black Holes and
  Entropy}},\ }\href {https://link.aps.org/doi/10.1103/PhysRevD.7.2333}
  {\bibfield  {journal} {\bibinfo  {journal} {Phys. Rev. D}\ }\textbf {\bibinfo
  {volume} {7}},\ \bibinfo {pages} {2333} (\bibinfo {year} {1973})}\BibitemShut
  {NoStop}%
\bibitem [{\citenamefont {Hawking}(1975)}]{hawking1975particle}%
  \BibitemOpen
  \bibfield  {author} {\bibinfo {author} {\bibfnamefont {S.~W.}\ \bibnamefont
  {Hawking}},\ }\bibfield  {title} {\bibinfo {title} {Particle creation by
  black holes},\ }\href {https://doi.org/10.1007/BF02345020} {\bibfield
  {journal} {\bibinfo  {journal} {Commun. Math. Phys.}\ }\textbf {\bibinfo
  {volume} {43}},\ \bibinfo {pages} {199} (\bibinfo {year} {1975})}\BibitemShut
  {NoStop}%
\bibitem [{\citenamefont {Ruppeiner}(1979)}]{ruppeiner1979thermodynamics}%
  \BibitemOpen
  \bibfield  {author} {\bibinfo {author} {\bibfnamefont {G.}~\bibnamefont
  {Ruppeiner}},\ }\bibfield  {title} {\bibinfo {title} {{Thermodynamics: A
  Riemannian geometric model}},\ }\href
  {https://link.aps.org/doi/10.1103/PhysRevA.20.1608} {\bibfield  {journal}
  {\bibinfo  {journal} {Phys. Rev. A}\ }\textbf {\bibinfo {volume} {20}},\
  \bibinfo {pages} {1608} (\bibinfo {year} {1979})}\BibitemShut {NoStop}%
\bibitem [{\citenamefont {Jacobson}(1995)}]{jacobson1995thermodynamics}%
  \BibitemOpen
  \bibfield  {author} {\bibinfo {author} {\bibfnamefont {T.}~\bibnamefont
  {Jacobson}},\ }\bibfield  {title} {\bibinfo {title} {{Thermodynamics of
  Spacetime: The Einstein Equation of State}},\ }\href
  {https://link.aps.org/doi/10.1103/PhysRevLett.75.1260} {\bibfield  {journal}
  {\bibinfo  {journal} {Phys. Rev. Lett.}\ }\textbf {\bibinfo {volume} {75}},\
  \bibinfo {pages} {1260} (\bibinfo {year} {1995})}\BibitemShut {NoStop}%
\bibitem [{\citenamefont {Padmanabhan}(2010)}]{padmanabhan2010thermodynamical}%
  \BibitemOpen
  \bibfield  {author} {\bibinfo {author} {\bibfnamefont {T.}~\bibnamefont
  {Padmanabhan}},\ }\bibfield  {title} {\bibinfo {title} {Thermodynamical
  aspects of gravity: new insights},\ }\href
  {https://iopscience.iop.org/article/10.1088/0034-4885/73/4/046901} {\bibfield
   {journal} {\bibinfo  {journal} {Rep. Prog. Phys.}\ }\textbf {\bibinfo
  {volume} {73}},\ \bibinfo {pages} {046901} (\bibinfo {year}
  {2010})}\BibitemShut {NoStop}%
\bibitem [{\citenamefont {Verlinde}(2011)}]{verlinde2011origin}%
  \BibitemOpen
  \bibfield  {author} {\bibinfo {author} {\bibfnamefont {E.}~\bibnamefont
  {Verlinde}},\ }\bibfield  {title} {\bibinfo {title} {{On the origin of
  gravity and the laws of Newton}},\ }\href
  {https://doi.org/10.1007/JHEP04(2011)029} {\bibfield  {journal} {\bibinfo
  {journal} {J. High Energy Phys.}\ }\textbf {\bibinfo {volume} {2011}}\bibinfo
   {number} { (4)},\ \bibinfo {pages} {1}}\BibitemShut {NoStop}%
\bibitem [{\citenamefont {Carroll}\ and\ \citenamefont
  {Remmen}(2016)}]{carroll2016what}%
  \BibitemOpen
\bibfield  {number} {  }\bibfield  {author} {\bibinfo {author} {\bibfnamefont
  {S.~M.}\ \bibnamefont {Carroll}}\ and\ \bibinfo {author} {\bibfnamefont
  {G.~N.}\ \bibnamefont {Remmen}},\ }\bibfield  {title} {\bibinfo {title} {What
  is the entropy in entropic gravity?},\ }\href
  {https://link.aps.org/doi/10.1103/PhysRevD.93.124052} {\bibfield  {journal}
  {\bibinfo  {journal} {Phys. Rev. D}\ }\textbf {\bibinfo {volume} {93}},\
  \bibinfo {pages} {124052} (\bibinfo {year} {2016})}\BibitemShut {NoStop}%
\bibitem [{\citenamefont {Bianconi}(2025)}]{bianconi2025gravity}%
  \BibitemOpen
  \bibfield  {author} {\bibinfo {author} {\bibfnamefont {G.}~\bibnamefont
  {Bianconi}},\ }\bibfield  {title} {\bibinfo {title} {Gravity from entropy},\
  }\href {https://link.aps.org/doi/10.1103/PhysRevD.111.066001} {\bibfield
  {journal} {\bibinfo  {journal} {Phys. Rev. D}\ }\textbf {\bibinfo {volume}
  {111}},\ \bibinfo {pages} {066001} (\bibinfo {year} {2025})}\BibitemShut
  {NoStop}%
\bibitem [{\citenamefont {Ahn}\ \emph {et~al.}(2022)\citenamefont {Ahn},
  \citenamefont {Guo}, \citenamefont {Nagaosa},\ and\ \citenamefont
  {Vishwanath}}]{ahn2022riemannian}%
  \BibitemOpen
  \bibfield  {author} {\bibinfo {author} {\bibfnamefont {J.}~\bibnamefont
  {Ahn}}, \bibinfo {author} {\bibfnamefont {G.-Y.}\ \bibnamefont {Guo}},
  \bibinfo {author} {\bibfnamefont {N.}~\bibnamefont {Nagaosa}},\ and\ \bibinfo
  {author} {\bibfnamefont {A.}~\bibnamefont {Vishwanath}},\ }\bibfield  {title}
  {\bibinfo {title} {Riemannian geometry of resonant optical responses},\
  }\href {https://doi.org/10.1038/s41567-021-01465-z} {\bibfield  {journal}
  {\bibinfo  {journal} {Nat. Phys.}\ }\textbf {\bibinfo {volume} {18}},\
  \bibinfo {pages} {290} (\bibinfo {year} {2022})}\BibitemShut {NoStop}%
\bibitem [{\citenamefont {Bouhon}\ \emph {et~al.}(2023)\citenamefont {Bouhon},
  \citenamefont {Timmel},\ and\ \citenamefont {Slager}}]{bouhon2023quantum}%
  \BibitemOpen
  \bibfield  {author} {\bibinfo {author} {\bibfnamefont {A.}~\bibnamefont
  {Bouhon}}, \bibinfo {author} {\bibfnamefont {A.}~\bibnamefont {Timmel}},\
  and\ \bibinfo {author} {\bibfnamefont {R.-J.}\ \bibnamefont {Slager}},\
  }\bibfield  {title} {\bibinfo {title} {Quantum geometry beyond projective
  single bands},\ }\href {https://arxiv.org/abs/2303.02180} {\bibfield
  {journal} {\bibinfo  {journal} {arXiv:2303.02180}\ } (\bibinfo {year}
  {2023})}\BibitemShut {NoStop}%
\bibitem [{\citenamefont {Mitscherling}\ \emph {et~al.}(2024)\citenamefont
  {Mitscherling}, \citenamefont {Avdoshkin},\ and\ \citenamefont
  {Moore}}]{mitscherling2024gauge}%
  \BibitemOpen
  \bibfield  {author} {\bibinfo {author} {\bibfnamefont {J.}~\bibnamefont
  {Mitscherling}}, \bibinfo {author} {\bibfnamefont {A.}~\bibnamefont
  {Avdoshkin}},\ and\ \bibinfo {author} {\bibfnamefont {J.~E.}\ \bibnamefont
  {Moore}},\ }\bibfield  {title} {\bibinfo {title} {Gauge-invariant projector
  calculus for quantum state geometry and applications to observables in
  crystals},\ }\href {https://arxiv.org/abs/2412.03637} {\bibfield  {journal}
  {\bibinfo  {journal} {arXiv:2412.03637}\ } (\bibinfo {year}
  {2024})}\BibitemShut {NoStop}%
\bibitem [{\citenamefont {Avdoshkin}\ \emph {et~al.}(2024)\citenamefont
  {Avdoshkin}, \citenamefont {Mitscherling},\ and\ \citenamefont
  {Moore}}]{avdoshkin2024multi}%
  \BibitemOpen
  \bibfield  {author} {\bibinfo {author} {\bibfnamefont {A.}~\bibnamefont
  {Avdoshkin}}, \bibinfo {author} {\bibfnamefont {J.}~\bibnamefont
  {Mitscherling}},\ and\ \bibinfo {author} {\bibfnamefont {J.~E.}\ \bibnamefont
  {Moore}},\ }\bibfield  {title} {\bibinfo {title} {The multi-state geometry of
  shift current and polarization},\ }\href {https://arxiv.org/abs/2409.16358}
  {\bibfield  {journal} {\bibinfo  {journal} {arXiv:2409.16358}\ } (\bibinfo
  {year} {2024})}\BibitemShut {NoStop}%
\bibitem [{\citenamefont {Jankowski}\ and\ \citenamefont
  {Slager}(2024)}]{jankowski2024quantized}%
  \BibitemOpen
  \bibfield  {author} {\bibinfo {author} {\bibfnamefont {W.~J.}\ \bibnamefont
  {Jankowski}}\ and\ \bibinfo {author} {\bibfnamefont {R.-J.}\ \bibnamefont
  {Slager}},\ }\bibfield  {title} {\bibinfo {title} {{Quantized Integrated
  Shift Effect in Multigap Topological Phases}},\ }\href
  {https://link.aps.org/doi/10.1103/PhysRevLett.133.186601} {\bibfield
  {journal} {\bibinfo  {journal} {Phys. Rev. Lett.}\ }\textbf {\bibinfo
  {volume} {133}},\ \bibinfo {pages} {186601} (\bibinfo {year}
  {2024})}\BibitemShut {NoStop}%
\bibitem [{\citenamefont {Jankowski}\ \emph
  {et~al.}(2025{\natexlab{b}})\citenamefont {Jankowski}, \citenamefont
  {Slager},\ and\ \citenamefont {Pizzochero}}]{jankowski2025enhancing}%
  \BibitemOpen
  \bibfield  {author} {\bibinfo {author} {\bibfnamefont {W.~J.}\ \bibnamefont
  {Jankowski}}, \bibinfo {author} {\bibfnamefont {R.-J.}\ \bibnamefont
  {Slager}},\ and\ \bibinfo {author} {\bibfnamefont {M.}~\bibnamefont
  {Pizzochero}},\ }\bibfield  {title} {\bibinfo {title} {Enhancing the
  hyperpolarizability of crystals with quantum geometry},\ }\href
  {https://arxiv.org/abs/2502.02660} {\bibfield  {journal} {\bibinfo  {journal}
  {arXiv:2502.02660}\ } (\bibinfo {year} {2025}{\natexlab{b}})}\BibitemShut
  {NoStop}%
\bibitem [{Note1()}]{Note1}%
  \BibitemOpen
  \bibinfo {note} {Here, we refer to an $n$-state quantum geometric object as
  one that carries $n$ band indices. For example, $g_{\mu \nu }^a$ and $g_{\mu
  \nu }^{ab}$ are the single- and two-state quantum metric tensors,
  respectively.}\BibitemShut {Stop}%
\bibitem [{\citenamefont {Misner}\ \emph {et~al.}(1973)\citenamefont {Misner},
  \citenamefont {Thorne},\ and\ \citenamefont
  {Wheeler}}]{misner1973gravitation}%
  \BibitemOpen
  \bibfield  {author} {\bibinfo {author} {\bibfnamefont {C.~W.}\ \bibnamefont
  {Misner}}, \bibinfo {author} {\bibfnamefont {K.~S.}\ \bibnamefont {Thorne}},\
  and\ \bibinfo {author} {\bibfnamefont {J.~A.}\ \bibnamefont {Wheeler}},\
  }\href@noop {} {\emph {\bibinfo {title} {Gravitation}}}\ (\bibinfo
  {publisher} {Macmillan},\ \bibinfo {year} {1973})\BibitemShut {NoStop}%
\bibitem [{\citenamefont {Wald}(2010)}]{wald2010general}%
  \BibitemOpen
  \bibfield  {author} {\bibinfo {author} {\bibfnamefont {R.~M.}\ \bibnamefont
  {Wald}},\ }\href@noop {} {\emph {\bibinfo {title} {General relativity}}}\
  (\bibinfo  {publisher} {University of Chicago press},\ \bibinfo {year}
  {2010})\BibitemShut {NoStop}%
\bibitem [{\citenamefont {Souza}\ \emph {et~al.}(2000)\citenamefont {Souza},
  \citenamefont {Wilkens},\ and\ \citenamefont
  {Martin}}]{souza2000polarization}%
  \BibitemOpen
  \bibfield  {author} {\bibinfo {author} {\bibfnamefont {I.}~\bibnamefont
  {Souza}}, \bibinfo {author} {\bibfnamefont {T.}~\bibnamefont {Wilkens}},\
  and\ \bibinfo {author} {\bibfnamefont {R.~M.}\ \bibnamefont {Martin}},\
  }\bibfield  {title} {\bibinfo {title} {{Polarization and localization in
  insulators: Generating function approach}},\ }\href
  {https://link.aps.org/doi/10.1103/PhysRevB.62.1666} {\bibfield  {journal}
  {\bibinfo  {journal} {Phys. Rev. B}\ }\textbf {\bibinfo {volume} {62}},\
  \bibinfo {pages} {1666} (\bibinfo {year} {2000})}\BibitemShut {NoStop}%
\bibitem [{\citenamefont {Michishita}\ and\ \citenamefont
  {Nagaosa}(2022)}]{Michishita2022dissipation}%
  \BibitemOpen
  \bibfield  {author} {\bibinfo {author} {\bibfnamefont {Y.}~\bibnamefont
  {Michishita}}\ and\ \bibinfo {author} {\bibfnamefont {N.}~\bibnamefont
  {Nagaosa}},\ }\bibfield  {title} {\bibinfo {title} {Dissipation and geometry
  in nonlinear quantum transports of multiband electronic systems},\ }\href
  {https://link.aps.org/doi/10.1103/PhysRevB.106.125114} {\bibfield  {journal}
  {\bibinfo  {journal} {Phys. Rev. B}\ }\textbf {\bibinfo {volume} {106}},\
  \bibinfo {pages} {125114} (\bibinfo {year} {2022})}\BibitemShut {NoStop}%
\bibitem [{\citenamefont {Chen}\ and\ \citenamefont {von
  Gersdorff}(2022)}]{chen2022measurement}%
  \BibitemOpen
  \bibfield  {author} {\bibinfo {author} {\bibfnamefont {W.}~\bibnamefont
  {Chen}}\ and\ \bibinfo {author} {\bibfnamefont {G.}~\bibnamefont {von
  Gersdorff}},\ }\bibfield  {title} {\bibinfo {title} {{Measurement of
  interaction-dressed Berry curvature and quantum metric in solids by optical
  absorption}},\ }\href {https://scipost.org/10.21468/SciPostPhysCore.5.3.040}
  {\bibfield  {journal} {\bibinfo  {journal} {SciPost Phys. Core}\ }\textbf
  {\bibinfo {volume} {5}},\ \bibinfo {pages} {040} (\bibinfo {year}
  {2022})}\BibitemShut {NoStop}%
\bibitem [{\citenamefont {Kashihara}\ \emph {et~al.}(2023)\citenamefont
  {Kashihara}, \citenamefont {Michishita},\ and\ \citenamefont
  {Peters}}]{kashihara2023quantum}%
  \BibitemOpen
  \bibfield  {author} {\bibinfo {author} {\bibfnamefont {T.}~\bibnamefont
  {Kashihara}}, \bibinfo {author} {\bibfnamefont {Y.}~\bibnamefont
  {Michishita}},\ and\ \bibinfo {author} {\bibfnamefont {R.}~\bibnamefont
  {Peters}},\ }\bibfield  {title} {\bibinfo {title} {{Quantum metric on the
  Brillouin zone in correlated electron systems and its relation to topology
  for Chern insulators}},\ }\href
  {https://link.aps.org/doi/10.1103/PhysRevB.107.125116} {\bibfield  {journal}
  {\bibinfo  {journal} {Phys. Rev. B}\ }\textbf {\bibinfo {volume} {107}},\
  \bibinfo {pages} {125116} (\bibinfo {year} {2023})}\BibitemShut {NoStop}%
\bibitem [{\citenamefont {Zhou}\ \emph {et~al.}(2024)\citenamefont {Zhou},
  \citenamefont {Hou}, \citenamefont {Wang}, \citenamefont {Tang},
  \citenamefont {Guo},\ and\ \citenamefont {Chien}}]{zhou2024sloqvist}%
  \BibitemOpen
  \bibfield  {author} {\bibinfo {author} {\bibfnamefont {Z.}~\bibnamefont
  {Zhou}}, \bibinfo {author} {\bibfnamefont {X.-Y.}\ \bibnamefont {Hou}},
  \bibinfo {author} {\bibfnamefont {X.}~\bibnamefont {Wang}}, \bibinfo {author}
  {\bibfnamefont {J.-C.}\ \bibnamefont {Tang}}, \bibinfo {author}
  {\bibfnamefont {H.}~\bibnamefont {Guo}},\ and\ \bibinfo {author}
  {\bibfnamefont {C.-C.}\ \bibnamefont {Chien}},\ }\bibfield  {title} {\bibinfo
  {title} {Sj\"oqvist quantum geometric tensor of finite-temperature mixed
  states},\ }\href {https://link.aps.org/doi/10.1103/PhysRevB.110.035404}
  {\bibfield  {journal} {\bibinfo  {journal} {Phys. Rev. B}\ }\textbf {\bibinfo
  {volume} {110}},\ \bibinfo {pages} {035404} (\bibinfo {year}
  {2024})}\BibitemShut {NoStop}%
\bibitem [{\citenamefont {Romeral}\ \emph {et~al.}(2025)\citenamefont
  {Romeral}, \citenamefont {Cummings},\ and\ \citenamefont
  {Roche}}]{romeral2025scaling}%
  \BibitemOpen
  \bibfield  {author} {\bibinfo {author} {\bibfnamefont {J.~M.}\ \bibnamefont
  {Romeral}}, \bibinfo {author} {\bibfnamefont {A.~W.}\ \bibnamefont
  {Cummings}},\ and\ \bibinfo {author} {\bibfnamefont {S.}~\bibnamefont
  {Roche}},\ }\bibfield  {title} {\bibinfo {title} {{Scaling of the integrated
  quantum metric in disordered topological phases}},\ }\href
  {https://link.aps.org/doi/10.1103/PhysRevB.111.134201} {\bibfield  {journal}
  {\bibinfo  {journal} {Phys. Rev. B}\ }\textbf {\bibinfo {volume} {111}},\
  \bibinfo {pages} {134201} (\bibinfo {year} {2025})}\BibitemShut {NoStop}%
\bibitem [{\citenamefont {Sukhachov}\ \emph {et~al.}(2025)\citenamefont
  {Sukhachov}, \citenamefont {Aase}, \citenamefont {M\ae{}land},\ and\
  \citenamefont {Sudb\o{}}}]{sukhachov2025effect}%
  \BibitemOpen
  \bibfield  {author} {\bibinfo {author} {\bibfnamefont {P.}~\bibnamefont
  {Sukhachov}}, \bibinfo {author} {\bibfnamefont {N.~H.}\ \bibnamefont {Aase}},
  \bibinfo {author} {\bibfnamefont {K.}~\bibnamefont {M\ae{}land}},\ and\
  \bibinfo {author} {\bibfnamefont {A.}~\bibnamefont {Sudb\o{}}},\ }\bibfield
  {title} {\bibinfo {title} {{Effect of the Hubbard interaction on the quantum
  metric}},\ }\href {https://link.aps.org/doi/10.1103/PhysRevB.111.085143}
  {\bibfield  {journal} {\bibinfo  {journal} {Phys. Rev. B}\ }\textbf {\bibinfo
  {volume} {111}},\ \bibinfo {pages} {085143} (\bibinfo {year}
  {2025})}\BibitemShut {NoStop}%
\bibitem [{\citenamefont {Blount}(1962)}]{blount1962formalisms}%
  \BibitemOpen
  \bibfield  {author} {\bibinfo {author} {\bibfnamefont {E.}~\bibnamefont
  {Blount}},\ }\bibfield  {title} {\bibinfo {title} {Formalisms of band
  theory},\ }in\ \href@noop {} {\emph {\bibinfo {booktitle} {Solid state
  physics}}},\ Vol.~\bibinfo {volume} {13}\ (\bibinfo  {publisher} {Elsevier},\
  \bibinfo {year} {1962})\ pp.\ \bibinfo {pages} {305--373}\BibitemShut
  {NoStop}%
\bibitem [{Note2()}]{Note2}%
  \BibitemOpen
  \bibinfo {note} {Regarding notation, we use Greek letters for real- or
  momentum-space indices--with Einstein summation implied--and Latin letters
  for band indices. Furthermore, $\protect \mathcal {O}_{(\mu \nu )} \equiv
  (\protect \mathcal {O}_{\mu \nu } + \protect \mathcal {O}_{\nu \mu })/2$ and
  $\protect \mathcal {O}_{[\mu \nu ]} \equiv (\protect \mathcal {O}_{\mu \nu }
  - \protect \mathcal {O}_{\nu \mu })/2$}\BibitemShut {NoStop}%
\bibitem [{\citenamefont {Ma}\ \emph {et~al.}(2010)\citenamefont {Ma},
  \citenamefont {Chen}, \citenamefont {Fan},\ and\ \citenamefont
  {Liu}}]{ma2010abelian}%
  \BibitemOpen
  \bibfield  {author} {\bibinfo {author} {\bibfnamefont {Y.-Q.}\ \bibnamefont
  {Ma}}, \bibinfo {author} {\bibfnamefont {S.}~\bibnamefont {Chen}}, \bibinfo
  {author} {\bibfnamefont {H.}~\bibnamefont {Fan}},\ and\ \bibinfo {author}
  {\bibfnamefont {W.-M.}\ \bibnamefont {Liu}},\ }\bibfield  {title} {\bibinfo
  {title} {{Abelian and non-Abelian quantum geometric tensor}},\ }\href
  {https://link.aps.org/doi/10.1103/PhysRevB.81.245129} {\bibfield  {journal}
  {\bibinfo  {journal} {Phys. Rev. B}\ }\textbf {\bibinfo {volume} {81}},\
  \bibinfo {pages} {245129} (\bibinfo {year} {2010})}\BibitemShut {NoStop}%
\bibitem [{\citenamefont {Atencia}\ \emph {et~al.}(2022)\citenamefont
  {Atencia}, \citenamefont {Niu},\ and\ \citenamefont
  {Culcer}}]{atencia2022semiclassical}%
  \BibitemOpen
  \bibfield  {author} {\bibinfo {author} {\bibfnamefont {R.~B.}\ \bibnamefont
  {Atencia}}, \bibinfo {author} {\bibfnamefont {Q.}~\bibnamefont {Niu}},\ and\
  \bibinfo {author} {\bibfnamefont {D.}~\bibnamefont {Culcer}},\ }\bibfield
  {title} {\bibinfo {title} {Semiclassical response of disordered conductors:
  Extrinsic carrier velocity and spin and field-corrected collision integral},\
  }\href {https://link.aps.org/doi/10.1103/PhysRevResearch.4.013001} {\bibfield
   {journal} {\bibinfo  {journal} {Phys. Rev. Res.}\ }\textbf {\bibinfo
  {volume} {4}},\ \bibinfo {pages} {013001} (\bibinfo {year}
  {2022})}\BibitemShut {NoStop}%
\bibitem [{\citenamefont {Mehraeen}(2024)}]{mehraeen2024quantum}%
  \BibitemOpen
  \bibfield  {author} {\bibinfo {author} {\bibfnamefont {M.}~\bibnamefont
  {Mehraeen}},\ }\bibfield  {title} {\bibinfo {title} {Quantum kinetic theory
  of quadratic responses},\ }\href
  {https://link.aps.org/doi/10.1103/PhysRevB.110.174423} {\bibfield  {journal}
  {\bibinfo  {journal} {Phys. Rev. B}\ }\textbf {\bibinfo {volume} {110}},\
  \bibinfo {pages} {174423} (\bibinfo {year} {2024})}\BibitemShut {NoStop}%
\bibitem [{\citenamefont {Parker}\ \emph {et~al.}(2019)\citenamefont {Parker},
  \citenamefont {Morimoto}, \citenamefont {Orenstein},\ and\ \citenamefont
  {Moore}}]{parker2019diagrammatic}%
  \BibitemOpen
  \bibfield  {author} {\bibinfo {author} {\bibfnamefont {D.~E.}\ \bibnamefont
  {Parker}}, \bibinfo {author} {\bibfnamefont {T.}~\bibnamefont {Morimoto}},
  \bibinfo {author} {\bibfnamefont {J.}~\bibnamefont {Orenstein}},\ and\
  \bibinfo {author} {\bibfnamefont {J.~E.}\ \bibnamefont {Moore}},\ }\bibfield
  {title} {\bibinfo {title} {{Diagrammatic approach to nonlinear optical
  response with application to Weyl semimetals}},\ }\href
  {https://link.aps.org/doi/10.1103/PhysRevB.99.045121} {\bibfield  {journal}
  {\bibinfo  {journal} {Phys. Rev. B}\ }\textbf {\bibinfo {volume} {99}},\
  \bibinfo {pages} {045121} (\bibinfo {year} {2019})}\BibitemShut {NoStop}%
\bibitem [{\citenamefont {Michishita}\ and\ \citenamefont
  {Peters}(2021)}]{Michishita2021effects}%
  \BibitemOpen
  \bibfield  {author} {\bibinfo {author} {\bibfnamefont {Y.}~\bibnamefont
  {Michishita}}\ and\ \bibinfo {author} {\bibfnamefont {R.}~\bibnamefont
  {Peters}},\ }\bibfield  {title} {\bibinfo {title} {{Effects of
  renormalization and non-Hermiticity on nonlinear responses in strongly
  correlated electron systems}},\ }\href
  {https://link.aps.org/doi/10.1103/PhysRevB.103.195133} {\bibfield  {journal}
  {\bibinfo  {journal} {Phys. Rev. B}\ }\textbf {\bibinfo {volume} {103}},\
  \bibinfo {pages} {195133} (\bibinfo {year} {2021})}\BibitemShut {NoStop}%
\bibitem [{Note3()}]{Note3}%
  \BibitemOpen
  \bibinfo {note} {See the Supplemental Material at [URL] for details on the
  evaluation of the response functions, equations of motion, renormalization
  functions and the dissipative EFE. This includes Refs.~\cite {mahan2000many,
  carroll2019spacetime}}\BibitemShut {NoStop}%
\bibitem [{Note4()}]{Note4}%
  \BibitemOpen
  \bibinfo {note} {We also note that this choice of dressing has the favorable
  property of preserving the $a \leftrightarrow b$ symmetries of the quantum
  metric and Berry curvature in band space.}\BibitemShut {Stop}%
\bibitem [{Note5()}]{Note5}%
  \BibitemOpen
  \bibinfo {note} {Note that we define the dressed three-state QGT in terms of
  the first two band indices as $\protect \tilde {Q}_{\mu \nu \rho }^{abc} =
  Q_{\mu \nu \rho }^{abc}/1+(\eta ^{ab})^2$}\BibitemShut {NoStop}%
\bibitem [{\citenamefont {Ziman}(2001)}]{ziman2001electrons}%
  \BibitemOpen
  \bibfield  {author} {\bibinfo {author} {\bibfnamefont {J.~M.}\ \bibnamefont
  {Ziman}},\ }\href@noop {} {\emph {\bibinfo {title} {Electrons and phonons:
  the theory of transport phenomena in solids}}}\ (\bibinfo  {publisher}
  {Oxford university press},\ \bibinfo {year} {2001})\BibitemShut {NoStop}%
\bibitem [{\citenamefont {Huang}\ \emph {et~al.}(2025)\citenamefont {Huang},
  \citenamefont {Xiao}, \citenamefont {Yang},\ and\ \citenamefont
  {Li}}]{huang2025scaling}%
  \BibitemOpen
  \bibfield  {author} {\bibinfo {author} {\bibfnamefont {Y.-X.}\ \bibnamefont
  {Huang}}, \bibinfo {author} {\bibfnamefont {C.}~\bibnamefont {Xiao}},
  \bibinfo {author} {\bibfnamefont {S.~A.}\ \bibnamefont {Yang}},\ and\
  \bibinfo {author} {\bibfnamefont {X.}~\bibnamefont {Li}},\ }\bibfield
  {title} {\bibinfo {title} {Scaling law and extrinsic mechanisms for
  time-reversal-odd second-order nonlinear transport},\ }\href
  {https://link.aps.org/doi/10.1103/PhysRevB.111.155127} {\bibfield  {journal}
  {\bibinfo  {journal} {Phys. Rev. B}\ }\textbf {\bibinfo {volume} {111}},\
  \bibinfo {pages} {155127} (\bibinfo {year} {2025})}\BibitemShut {NoStop}%
\bibitem [{\citenamefont {Liu}\ \emph {et~al.}(2024{\natexlab{b}})\citenamefont
  {Liu}, \citenamefont {Zhang}, \citenamefont {Zhu},\ and\ \citenamefont
  {Su}}]{liu2024effect}%
  \BibitemOpen
  \bibfield  {author} {\bibinfo {author} {\bibfnamefont {Z.}~\bibnamefont
  {Liu}}, \bibinfo {author} {\bibfnamefont {Z.-F.}\ \bibnamefont {Zhang}},
  \bibinfo {author} {\bibfnamefont {Z.-G.}\ \bibnamefont {Zhu}},\ and\ \bibinfo
  {author} {\bibfnamefont {G.}~\bibnamefont {Su}},\ }\bibfield  {title}
  {\bibinfo {title} {{Effect of disorder on Berry curvature and quantum metric
  in two-band gapped graphene}},\ }\href
  {https://link.aps.org/doi/10.1103/PhysRevB.110.245419} {\bibfield  {journal}
  {\bibinfo  {journal} {Phys. Rev. B}\ }\textbf {\bibinfo {volume} {110}},\
  \bibinfo {pages} {245419} (\bibinfo {year} {2024}{\natexlab{b}})}\BibitemShut
  {NoStop}%
\bibitem [{\citenamefont {Gong}\ \emph {et~al.}(2024)\citenamefont {Gong},
  \citenamefont {Du}, \citenamefont {Sun}, \citenamefont {Lu},\ and\
  \citenamefont {Xie}}]{gong2024nonlinear}%
  \BibitemOpen
  \bibfield  {author} {\bibinfo {author} {\bibfnamefont {Z.-H.}\ \bibnamefont
  {Gong}}, \bibinfo {author} {\bibfnamefont {Z.}~\bibnamefont {Du}}, \bibinfo
  {author} {\bibfnamefont {H.-P.}\ \bibnamefont {Sun}}, \bibinfo {author}
  {\bibfnamefont {H.-Z.}\ \bibnamefont {Lu}},\ and\ \bibinfo {author}
  {\bibfnamefont {X.}~\bibnamefont {Xie}},\ }\bibfield  {title} {\bibinfo
  {title} {{Nonlinear transport theory at the order of quantum metric}},\
  }\href {https://arxiv.org/abs/2410.04995} {\bibfield  {journal} {\bibinfo
  {journal} {arXiv:2410.04995}\ } (\bibinfo {year} {2024})}\BibitemShut
  {NoStop}%
\bibitem [{\citenamefont {Konopik}\ and\ \citenamefont
  {Lutz}(2019)}]{konopik2019quantum}%
  \BibitemOpen
  \bibfield  {author} {\bibinfo {author} {\bibfnamefont {M.}~\bibnamefont
  {Konopik}}\ and\ \bibinfo {author} {\bibfnamefont {E.}~\bibnamefont {Lutz}},\
  }\bibfield  {title} {\bibinfo {title} {Quantum response theory for
  nonequilibrium steady states},\ }\href
  {https://link.aps.org/doi/10.1103/PhysRevResearch.1.033156} {\bibfield
  {journal} {\bibinfo  {journal} {Phys. Rev. Res.}\ }\textbf {\bibinfo {volume}
  {1}},\ \bibinfo {pages} {033156} (\bibinfo {year} {2019})}\BibitemShut
  {NoStop}%
\bibitem [{\citenamefont {Facchi}\ \emph {et~al.}(2010)\citenamefont {Facchi},
  \citenamefont {Kulkarni}, \citenamefont {Man'Ko}, \citenamefont {Marmo},
  \citenamefont {Sudarshan},\ and\ \citenamefont
  {Ventriglia}}]{facchi2010classical}%
  \BibitemOpen
  \bibfield  {author} {\bibinfo {author} {\bibfnamefont {P.}~\bibnamefont
  {Facchi}}, \bibinfo {author} {\bibfnamefont {R.}~\bibnamefont {Kulkarni}},
  \bibinfo {author} {\bibfnamefont {V.}~\bibnamefont {Man'Ko}}, \bibinfo
  {author} {\bibfnamefont {G.}~\bibnamefont {Marmo}}, \bibinfo {author}
  {\bibfnamefont {E.}~\bibnamefont {Sudarshan}},\ and\ \bibinfo {author}
  {\bibfnamefont {F.}~\bibnamefont {Ventriglia}},\ }\bibfield  {title}
  {\bibinfo {title} {{Classical and quantum Fisher information in the
  geometrical formulation of quantum mechanics}},\ }\href
  {https://doi.org/10.1016/j.physleta.2010.10.005} {\bibfield  {journal}
  {\bibinfo  {journal} {Phys. Lett. A}\ }\textbf {\bibinfo {volume} {374}},\
  \bibinfo {pages} {4801} (\bibinfo {year} {2010})}\BibitemShut {NoStop}%
\bibitem [{\citenamefont {Loaiza}\ \emph {et~al.}(2024)\citenamefont {Loaiza},
  \citenamefont {Motlagh}, \citenamefont {Hejazi}, \citenamefont {Zini},
  \citenamefont {Delgado},\ and\ \citenamefont
  {Arrazola}}]{loaiza2024nonlinear}%
  \BibitemOpen
  \bibfield  {author} {\bibinfo {author} {\bibfnamefont {I.}~\bibnamefont
  {Loaiza}}, \bibinfo {author} {\bibfnamefont {D.}~\bibnamefont {Motlagh}},
  \bibinfo {author} {\bibfnamefont {K.}~\bibnamefont {Hejazi}}, \bibinfo
  {author} {\bibfnamefont {M.~S.}\ \bibnamefont {Zini}}, \bibinfo {author}
  {\bibfnamefont {A.}~\bibnamefont {Delgado}},\ and\ \bibinfo {author}
  {\bibfnamefont {J.~M.}\ \bibnamefont {Arrazola}},\ }\bibfield  {title}
  {\bibinfo {title} {Nonlinear spectroscopy via generalized quantum phase
  estimation},\ }\href {https://arxiv.org/abs/2405.13885} {\bibfield  {journal}
  {\bibinfo  {journal} {arXiv:2405.13885}\ } (\bibinfo {year}
  {2024})}\BibitemShut {NoStop}%
\bibitem [{\citenamefont {Mahan}(2000)}]{mahan2000many}%
  \BibitemOpen
  \bibfield  {author} {\bibinfo {author} {\bibfnamefont {G.~D.}\ \bibnamefont
  {Mahan}},\ }\href@noop {} {\emph {\bibinfo {title} {\textit{Many Particle
  Physics, Third Edition}}}}\ (\bibinfo  {publisher} {Plenum},\ \bibinfo
  {address} {New York},\ \bibinfo {year} {2000})\BibitemShut {NoStop}%
\bibitem [{\citenamefont {Carroll}(2019)}]{carroll2019spacetime}%
  \BibitemOpen
  \bibfield  {author} {\bibinfo {author} {\bibfnamefont {S.~M.}\ \bibnamefont
  {Carroll}},\ }\href@noop {} {\emph {\bibinfo {title} {Spacetime and
  geometry}}}\ (\bibinfo  {publisher} {Cambridge University Press},\ \bibinfo
  {year} {2019})\BibitemShut {NoStop}%
\end{thebibliography}%

\clearpage
\onecolumngrid
\setcounter{equation}{0}
\setcounter{figure}{0}
\setcounter{table}{0}
\setcounter{page}{1}
\makeatletter
\renewcommand*{\thesection}{S\arabic{section}}
\renewcommand{\theequation}{S.\arabic{equation}}
\renewcommand{\thefigure}{S.\arabic{figure}}
\renewcommand*{\theHequation}{S.\arabic{equation}}
\renewcommand*{\theHfigure}{S.\arabic{figure}}
\renewcommand*{\theHtable}{S.\arabic{table}}
\renewcommand*{\theHsection}{S.\arabic{section}}
\makeatother

\begin{center}
\textbf{\large 
\medskip
Supplemental Material for ``Quantum response theory and momentum-space gravity"}
\end{center}

\section{S1.~Geometric origin of dissipative responses and dynamics}

In this section, after elaborating on the basic diagrammatic framework for evaluating conductivities in the presence of finite dissipation, we take the dc limit and derive the equation of motion that arises from the response functions. In doing so, we also show how both linear and nonlinear responses can be fully reexpressed in the geometric language, which, to our knowledge, has not been demonstrated before. Indeed, while there are several geometric presentations of \textit{specific} quadratic responses in the literature, a complete geometric classification of quadratic response theory that takes into account all Feynman diagrams simultaneously is yet to be done. In the derivations of the nonlinear responses presented below, several known elements of two-state quantum geometry make an appearance, including the quantum metric, Berry curvature and quantum connection. Furthermore, while two-state quantum geometry is certainly a necessary element, we find that it is not sufficient, as there are terms in the quadratic response functions which do not take closed two-state forms. This should perhaps not come as a surprise, as it would seem reasonable to question the complete classification of the three-point functions of quadratic response theory (involving traces over three distinct states) via two-state objects. As it turns out, we find that it is precisely the three-state QGT that is the missing necessary ingredient to complete the geometric classification.

\subsection{Response functions and frequency summations}

\begin{figure}[hpt]
\captionsetup[subfigure]{labelformat=empty}
    \sidesubfloat[]{\includegraphics[width=0.51\linewidth,trim={1cm 1cm 1cm .5cm}]{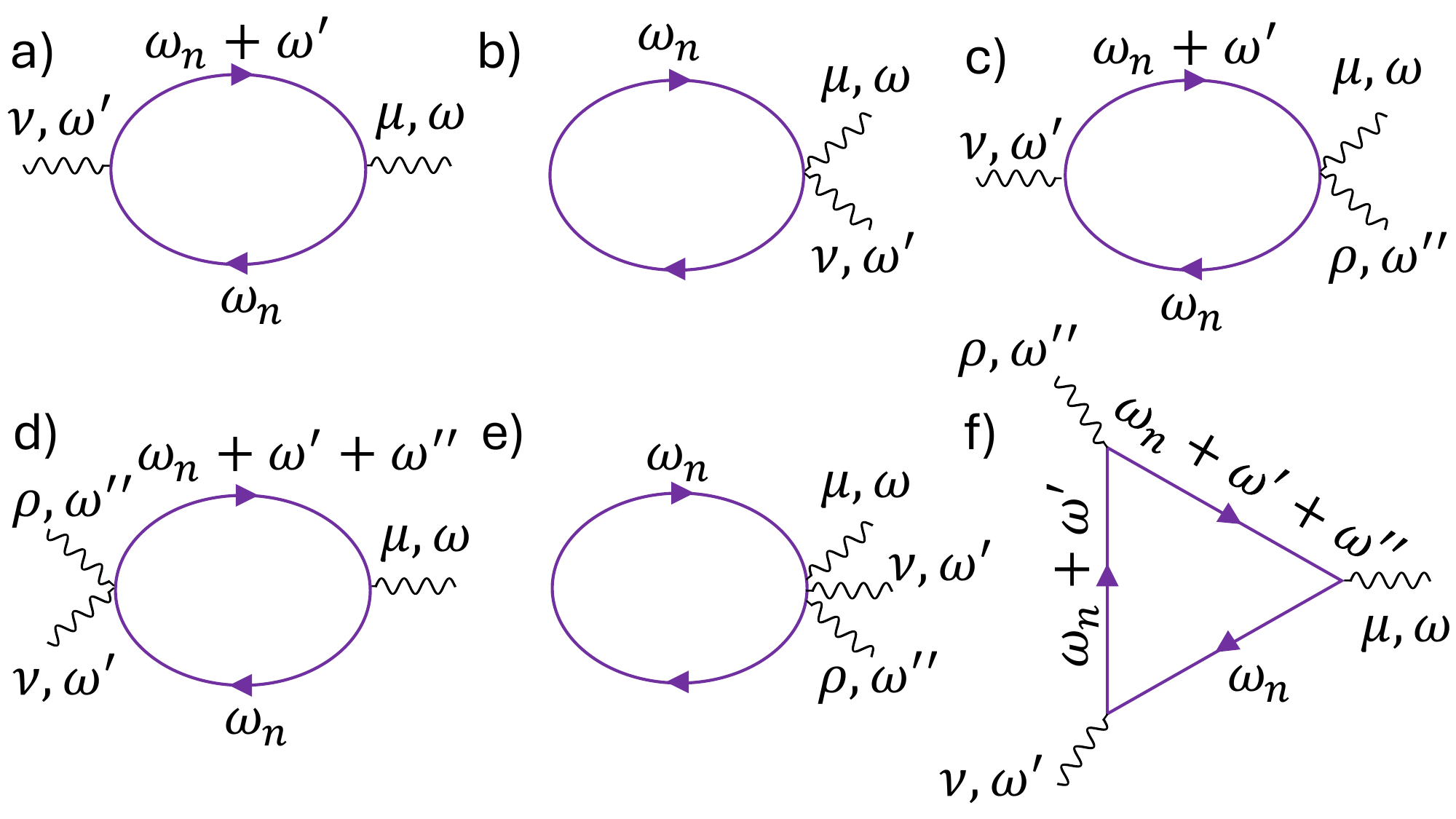}\label{fig1a}}
    \caption{Linear [(a) and (b)] and quadratic [(c)-(f)] conductivity diagrams. Loops and legs are electron and photon propagators, respectively, and vertices imply velocity operator insertions, with the index $\mu$ reserved for the output. The Feynman rules resulting in Eqs.~(\ref{eq_sigma1}) and (\ref{eq_sigma2}) consist of the following factors and procedures: 1) A factor of $1/k!$ for every set of $k$ connected photons. 2) Factors of $e/\hbar$ for the output photon and $ie/\chi$ for each input photon, with $\chi=\omega^{\pr},\omega^{\dpr}$. 3) Trace over momentum and band indices of the loop. 4) Matsubara frequency summation $1/\beta \sum_n$.}
    \label{figS1}
\end{figure}

Using a diagrammatic approach within the Matsubara formalism~\cite{parker2019diagrammatic}, and applying the Feynman rules presented in Fig.~{\ref{figS1}}, the linear and quadratic ac  conductivities are obtained as
\begin{equation}
\label{eq_sigma1}
\begin{split}
\sigma_{\mu \nu} (\omega; \omega^{\pr})
=&
\frac{ie^2}{\hbar \omega^{\pr}}\frac{1}{\beta}\sum_n
\text{Tr} \left[
h_{\mu} \mathcal{G}(\omega_n) 
h_{\nu} \mathcal{G}(\omega_n + \omega^{\pr})
+
h_{\mu \nu} \mathcal{G}(\omega_n)
\right],
\end{split}
\end{equation}
and
\begin{equation}
\label{eq_sigma2}
\begin{split}
\sigma_{\mu \nu \rho}
(\omega; \omega^{\pr}, \omega^{\dpr})
=&
\frac{e}{2 \hbar}
\frac{(ie)^2}{\omega^{\pr} \omega^{\dpr}}\frac{1}{\beta}\sum_n
\text{Tr} \left[
h_{\mu \rho} \mathcal{G}(\omega_n) 
h_{\nu} \mathcal{G}(\omega_n + \omega^{\pr})
+
\frac{1}{2} h_{\mu} \mathcal{G}(\omega_n) 
h_{\nu \rho} \mathcal{G}( \omega_n + \omega^{\pr} +\omega^{\dpr})
\right.
\\
&+
\left.
\frac{1}{2} h_{\mu \nu \rho} \mathcal{G}(\omega_n)
+
h_{\mu} \mathcal{G}(\omega_n) 
h_{\nu} \mathcal{G}(\omega_n + \omega^{\pr})
h_{\rho} \mathcal{G}(\omega_n + \omega^{\pr}
+ \omega^{\dpr})
+ \{(\nu, \omega^{\pr}) \leftrightarrow
(\rho, \omega^{\dpr}) \}
\right],
\end{split}
\end{equation}
where $\omega$ is the frequency of the output photon and $\omega^{\pr}, \omega^{\dpr}$ are input frequencies. The Matsubara Green's function is given by
$\mathcal{G}(\omega_n)
=
(i\omega_n - H - \Sigma)^{-1}$, where $H$ is the system Hamiltonian, with
$H \ket{\psi^a} = \veps^a \ket{\psi^a}$. And for the self-energy, we make the phenomenelogical approximation $\Sigma = -i\gamma/2$, with $\gamma$ measuring the strength of dissipation induced by scattering events. Finally, velocity operators are obtained by applying successive Berry covariant derivatives,
$h_{\mu_1 \ldots \mu_n}
=
\mathcal{D}_{\mu_1 \ldots \mu_n} H$.

To evaluate Eqs.~(\ref{eq_sigma1}) and (\ref{eq_sigma2}), the effect of the dissipation parameter $\gamma$ is included through a phenomenological shift of the frequency factors in the complex plane,
$\omega \rightarrow \omega + i\gamma$~\cite{parker2019diagrammatic}. And the Matsubara Green's function is expressed in the spectral representation as
\begin{equation}
\mathcal{G}^a (\omega_n)
=
\int_{-\infty}^{\infty} \frac{d\veps}{2\pi}
\frac{A^a(\veps)}{i\omega_n - \veps},
\end{equation}
where the denominator is the bare Green's function,
$\mathcal{G}_0^a (\omega_n)= (i \omega_n - \veps^a)^{-1}$ and
$A^a (\veps)= -2 \text{Im} G_R^a (\veps)$
is the spectral function~\cite{mahan2000many}, with $G_R$ the retarded Green's function. The frequency summations that are relevant to the evaluation of the conductivities are given by~\cite{mahan2000many, Michishita2021effects}
\begin{subequations}
\begin{align}
\frac{1}{\beta} \sum_n
\mathcal{G}_0^a (\omega_n)
&=
f^a,
\\
\frac{1}{\beta} \sum_n
\mathcal{G}_0^a (\omega_n)
\mathcal{G}_0^b (\omega_n+ \omega^{\pr})
&=
\frac{f^a - f^b}{\omega^{\pr} + \veps^{ab}},
\\
\frac{1}{\beta} \sum_n
\mathcal{G}_0^a (\omega_n)
\mathcal{G}_0^b (\omega_n+ \omega^{\pr})
\mathcal{G}_0^c (\omega_n+ \omega^{\pr}
+ \omega^{\dpr})
&=
\frac{
(\omega^{\dpr} - \veps^{cb})(f^a - f^b)
+
(\omega^{\pr} - \veps^{ba})(f^c - f^b)
}
{(\omega^{\pr} - \veps^{ba})
(\omega^{\dpr} - \veps^{cb})
(\omega^{\pr} + \omega^{\dpr} - \veps^{ca})},
\end{align}
\end{subequations}
with the shorthand $f^a \equiv f(\veps^a)$ for the Fermi distribution. In addition, an approximation is needed to obtain analytical results in the presence of dissipation. Taking
$\gamma \beta \ll1$
results in the identity~\cite{Michishita2021effects}
\begin{equation}
\int_{-\infty}^{\infty}
\frac{d\veps}{2 \pi}
A^a(\veps) 
F \left( \veps, \{\omega_n\} \right) f(\veps)
\simeq
F \left( \veps^a \pm \frac{i \gamma}{2}, 
\{\omega_n\} \right) f (\veps^a \pm \frac{i \gamma}{2} ),
\end{equation}
where $F$ is a general function that includes velocity operator components and the sign is chosen $\pm$ when $F$ is analytic in the upper/lower complex plane of the energy integrand. We note that this approximation is not as restricting as may seem, as values as large as $\gamma \beta=0.5$ have been shown to yield good agreement with numerical results as far as the conductivity is concerned~\cite{Michishita2021effects}. We apply these steps to Eqs.~(\ref{eq_sigma1}) and (\ref{eq_sigma2}) and take the dc limit
$\omega, \omega^{\pr}, \omega^{\dpr} \rightarrow 0$, yielding the frequency summations
\begin{subequations}
\label{freq_sum_dress}
\begin{align}
&\frac{1}{\beta} \sum_n
\mathcal{G}^a (\omega_n)
\mathcal{G}^b (\omega_n+ \omega^{\pr})
\rightarrow
\frac{ f (\veps^a + \frac{i \gamma}{2})
- f (\veps^b - \frac{i \gamma}{2})}
{\veps^{ab} + i \gamma}
\simeq
\frac{f^a - f^b}{\veps^{ab} + i\gamma}
+
i \frac{\gamma}{2}
\left(
\frac{\pd f^a}{\pd \veps^a} + \frac{\pd f^b}{\pd \veps^b}
\right)
\frac{1}{\veps^{ab} + i\gamma},
\\
&\frac{1}{\beta} \sum_n
\mathcal{G}^a (\omega_n)
\mathcal{G}^b (\omega_n+ \omega^{\pr})
\mathcal{G}^c (\omega_n+ \omega^{\pr}
+ \omega^{\dpr})
\rightarrow
\frac{ f (\veps^a + \frac{i \gamma}{2})
- f (\veps^b - \frac{i \gamma}{2})}
{(\veps^{ba} - i \gamma) (\veps^{ca} - i \gamma)}
+
\frac{ f (\veps^c - \frac{i \gamma}{2})
- f (\veps^b + \frac{i \gamma}{2})}
{(\veps^{cb} - i \gamma) (\veps^{ca} - i \gamma)}
\\
&\simeq
\frac{f^a - f^b}
{(\veps^{ba} - i\gamma) (\veps^{ca} - i\gamma)}
+
\frac{f^c - f^b}
{(\veps^{cb} - i\gamma) (\veps^{ca} - i\gamma)}
+
i \frac{\gamma}{2}
\left(
\frac{\pd f^a}{\pd \veps^a} + \frac{\pd f^b}{\pd \veps^b}
\right)
\frac{1}{(\veps^{ba} - i\gamma) (\veps^{ca} - i\gamma)}
\\
&\;\;\;\;-
i \frac{\gamma}{2}
\left(
\frac{\pd f^c}{\pd \veps^c} + \frac{\pd f^b}{\pd \veps^b}
\right)
\frac{1}{(\veps^{cb} - i\gamma) (\veps^{ca} - i\gamma)}.
\end{align}
\end{subequations}

\subsection{DC conductivities and dressed equation of motion}

Using the explanations of the previous section, the dc response functions can be evaluated. From Eq.~(\ref{freq_sum_dress}), it is clear that both the distribution function and its derivatives appear in the responses. As is shown below, for the purpose of deriving the equation of motion at nonlinear order, it is necessary to include the latter terms in order to obtain the correct signs of the carrier densities in the nonlinear response functions. 

The linear dc conductivity reads
\begin{equation}
\label{sig_munu_expanded}
\begin{split}
\sigma_{\mu\nu}
&=
\frac{e^2}{\hbar \gamma} \sum_{\mbk a}
\left(
f^a h_{\mu\nu}^a
+
\sum_b  \frac{f^a - f^b}{\veps^{ab} + i \gamma}
h_{\nu}^{ab} h_{\mu}^{ba}
\right)
\\
&=
\frac{e^2}{\gamma} \sum_{\mbk a}
f^a \pd_{\nu} v_{\mu}^a
+
\frac{i e^2}{\hbar} \sum_{\mbk ab}
(f^a - f^b) \frac{Q_{\mu\nu}^{ab}}{1 - i \eta^{ab}}
\\
&=
- \frac{e^2}{\gamma} \sum_{\mbk a}
\pd_{\nu} f^a v_{\mu}^a
+
\frac{e^2}{\hbar} \sum_{\mbk ab}
f^a \tilde{\Omega}_{\mu\nu}^{ab}
-
\frac{2 e^2}{\hbar} \sum_{\mbk ab}
f^a \eta^{ab} \tilde{g}_{\mu\nu}^{ab},
\end{split}
\end{equation}
yielding the linear current density
\begin{equation}
\frac{j_{\mu}^1}{-e}
=
\sum_{\mbk a}
n_1^a v_{\mu}^a
+
\dot{k}^{\nu} \sum_{\mbk ab}
f^a \tilde{\Omega}_{\mu\nu}^{ab}
-
2 \dot{k}^{\nu} \sum_{\mbk ab}
f^a \eta^{ab} \tilde{g}_{\mu\nu}^{ab},
\end{equation}
with the linear carrier density given by $n_1^a = - \tau \dot{k}^{\nu} \pd_{\nu} f^a$ and $\tau = \hbar/ \gamma$ the relaxation time. Note that Eq.~(\ref{sig_munu_expanded}) reveals the natural choice of dressing for the quantum geometric quantities in the sense that the Weyl-transformed Berry curvature and quantum metric represent the effective geometry the carriers experience in the presence of dissipation within the relaxation time approximation. This is then consistently applied in the evaluation of the nonlinear responses as well, the calculation of which is considerably more detailed. However, the contributions can be systematically analyzed and classified as follows. Taking into account the four terms in Eq.~(\ref{eq_sigma2}), we decompose the nonlinear conductivity tensor as
\begin{subequations}
\begin{align}
\sigma_{\mu\nu\rho}
&=
\sigma_{\mu\nu\rho}^\text{A}
+
\sigma_{\mu\nu\rho}^\text{B},
\\
\sigma_{\mu\nu\rho}^\text{A}
&=
\sigma_{\mu\nu\rho}^\text{I,A}
+
\sigma_{\mu\nu\rho}^\text{II,A}
+
\sigma_{\mu\nu\rho}^\text{III,A}
+
\sigma_{\mu\nu\rho}^\text{IV,A},
\\
\sigma_{\mu\nu\rho}^\text{B}
&=
\sigma_{\mu\nu\rho}^\text{I,B}
+
\sigma_{\mu\nu\rho}^\text{II,B}
+
\sigma_{\mu\nu\rho}^\text{III,B}
+
\sigma_{\mu\nu\rho}^\text{IV,B},
\end{align}
\end{subequations}
where the labels ``A" and ``B" specify Fermi distribution terms and Fermi distribution derivative terms, respectively, as per Eq.~(\ref{freq_sum_dress}). The former terms yield the contributions
\begin{equation}
\begin{split}
\sigma_{\mu\nu\rho}^\text{I,A}
&=
\frac{e^3}{2 \hbar \gamma^2}
\sum_{\mbk a}
f^a h_{\mu \nu \rho}^a
+ (\nu \leftrightarrow \rho)
=
\frac{e^3}{2 \hbar \gamma^2} \sum_{\mbk a} f^a
\left[
\hbar \pd_{\nu} \pd_{\rho} v_{\mu}^a
-
2 \sum_b \pd_{\mu} (g_{\mu\nu}^{ab} \veps^{ab})
\right]
\\
&-
\frac{i e^3}{2 \hbar \gamma^2} \sum_{\mbk ab} 
(f^a - f^b) \mathcal{A}_{\mu}^{\pr ab}
\left[
-i \pd_{\nu} (\mathcal{A}_{\rho}^{\pr ba} \veps^{ab})
-
i \mathcal{A}_{\nu}^{\pr ba} \pd_{\rho} \veps^{ab}
-
\sum_c (\mathcal{A}_{\nu}^{bc} \mathcal{A}_{\rho}^{ca} \veps^{ac}
-
\mathcal{A}_{\nu}^{ca} \mathcal{A}_{\rho}^{bc} \veps^{cb})
\right],
\end{split}
\end{equation}
\begin{equation}
\begin{split}
\sigma_{\mu\nu\rho}^\text{II,A}
&=
\frac{e^3}{\hbar \gamma^2} \sum_{\mbk ab}
\frac{f^a - f^b}{\veps^{ab} - i \gamma}
h_{\mu\rho}^{ab} h_{\nu}^{ba}
+ (\nu \leftrightarrow \rho)
\\
&=
\frac{e^3}{\hbar \gamma^2} \sum_{\mbk ab} 
\frac{f^a - f^b}{\veps^{ab} - i \gamma}
\mathcal{A}_{\nu}^{\pr ba} \veps^{ab}
\left[
\pd_{\mu} (\mathcal{A}_{\rho}^{\pr ab} \veps^{ab})
+
\mathcal{A}_{\mu}^{\pr ab} \pd_{\rho} \veps^{ab}
+
i \sum_c (\mathcal{A}_{\mu}^{ac} \mathcal{A}_{\rho}^{cb} \veps^{bc}
-
\mathcal{A}_{\rho}^{ac} \mathcal{A}_{\mu}^{cb} \veps^{ca})
\right],
\end{split}
\end{equation}
\begin{equation}
\begin{split}
\sigma_{\mu\nu\rho}^\text{III,A}
&=
\frac{e^3}{2 \hbar \gamma^2} \sum_{\mbk ab}
\frac{f^a - f^b}{\veps^{ab} + i \gamma}
h_{\nu\rho}^{ab} h_{\mu}^{ba}
+ (\nu \leftrightarrow \rho)
\\
&=
\frac{e^3}{2 \hbar \gamma^2} \sum_{\mbk ab} 
\frac{f^a - f^b}{\veps^{ab} + i \gamma}
\mathcal{A}_{\mu}^{\pr ba} \veps^{ab}
\left[
\pd_{\nu} (\mathcal{A}_{\rho}^{\pr ab} \veps^{ab})
+
\mathcal{A}_{\nu}^{\pr ab} \pd_{\rho} \veps^{ab}
+
i \sum_c (\mathcal{A}_{\nu}^{ac} \mathcal{A}_{\rho}^{cb} \veps^{bc}
-
\mathcal{A}_{\rho}^{ac} \mathcal{A}_{\nu}^{cb} \veps^{ca})
\right],
\end{split}
\end{equation}
\begin{equation}
\begin{split}
\sigma_{\mu\nu\rho}^\text{IV,A}
=&
\frac{e^3}{\hbar \gamma^2} \sum_{\mbk abc}
\left[
\frac{f^a - f^c}
{(\veps^{ac} - i \gamma) (\veps^{ab} - i \gamma)}
+
\frac{f^b - f^c}
{(\veps^{cb} - i \gamma) (\veps^{ab} - i \gamma)}
\right]
h_{\mu}^{ab} h_{\nu}^{bc} h_{\rho}^{ca}
+ (\nu \leftrightarrow \rho)
\\
=&
\frac{e^3}{\hbar \gamma^2} \sum_{\mbk abc}
\left[
\frac{f^a - f^c}
{(\veps^{ac} - i \gamma) (\veps^{ab} - i \gamma)}
+
\frac{f^b - f^c}
{(\veps^{cb} - i \gamma) (\veps^{ab} - i \gamma)}
\right]
\\
&\times
\left[
Q_{\nu\rho}^{ac} \delta^{ab} (\veps^{ac})^2 \pd_{\mu} \veps^a
+
Q_{\mu\rho}^{ab} \delta^{bc} (\veps^{ab})^2 \pd_{\nu} \veps^b
+
Q_{\mu\nu}^{ab} \delta^{ac} (\veps^{ab})^2 \pd_{\rho} \veps^a
-
i Q_{\mu\nu\rho}^{abc} \veps^{ab} \veps^{bc} \veps^{ca}
\right].
\end{split}
\end{equation}
Combining the four terms and after a little algebra, we arrive at the expression
\begin{equation}
\label{pre_sig2_A}
\begin{split}
\sigma_{\mu\nu\rho}^\text{A}
=&
\frac{e^3}{2 \gamma^2}
\sum_{\mbk a} f^a \pd_{\nu} \pd_{\rho} v_{\mu}^a
-
\frac{e^3}{2 \hbar \gamma} \sum_{\mbk a b} f^a \tilde{\Omega}_{\mu\nu}^{ab} \frac{\pd_{\rho} \veps^{ab}}{\veps^{ab}}
\frac{1- 3 (\eta^{ab})^2}{1 + (\eta^{ab})^2}
-
\frac{e^3}{\hbar} \sum_{\mbk a b} f^a \tilde{g}_{\mu\nu}^{ab} \pd_{\rho} \left( \frac{1}{\veps^{ab}} \right)
\frac{3- (\eta^{ab})^2}{1 + (\eta^{ab})^2}
\\
&+
\frac{e^3}{\hbar} \sum_{\mbk a b} f^a
\tilde{g}_{\nu \rho}^{ab} \pd_{\mu}
\left( \frac{1}{\veps^{ab}} \right)
\frac{1- (\eta^{ab})^2}{1 + (\eta^{ab})^2}
-
\frac{e^3}{\hbar \gamma^2}
\sum_{\mbk a b} f^a
\tilde{g}_{\nu \rho}^{ab} \pd_{\mu} \veps^{ab}
-
\frac{2 e^3}{\hbar} \sum_{\mbk a b} f^a
\frac{1}{\veps^{ab}}
\tilde{\Gamma}_{\nu\rho\mu}^{ab}
+
\sum_{i=1}^4 \mathcal{F}_i
\end{split}
\end{equation}
where
\begin{equation}
\begin{split}
\mathcal{F}_1
=&
- \frac{e^3}{\hbar \gamma} \sum_{\mbk a b} f^a
\text{Im} \bigg\{
\frac{\mathcal{A}_{\mu}^{\pr ab} 
(\mathcal{D}_{\nu} \left[ \mathcal{A}_{\rho}^{\pr} , H \right])^{ba}}
{\veps^{ab} - i \gamma} \bigg\}
\\
=&
-\frac{e^3}{\hbar \gamma} \sum_{\mbk a b} f^a
\frac{\veps^{ab}}{(\veps^{ab})^2 + \gamma^2}
\left[
- \frac{1}{2} \Omega_{\mu\rho}^{ab} \pd_{\nu} \veps^{ab}
+
\text{Im} (C_{\mu\nu\rho}^{ba}) \veps^{ab}
-
\sum_c \text{Re} (S_{\mu\nu\rho}^{abc}) (\veps^{bc} - \veps^{ca})
\right]
\\
&-
\frac{e^3}{\hbar} \sum_{\mbk a b} f^a
\frac{1}{(\veps^{ab})^2 + \gamma^2}
\left[
g_{\mu\rho}^{ab} \pd_{\nu} \veps^{ab}
+
\text{Re} (C_{\mu\nu\rho}^{ba}) \veps^{ab}
+
\sum_c \text{Im} (S_{\mu\nu\rho}^{abc}) (\veps^{bc} - \veps^{ca})
\right],
\end{split}
\end{equation}
\begin{equation}
\begin{split}
\mathcal{F}_2
=&
\frac{2 e^3}{\hbar \gamma} \sum_{\mbk a b} f^a
\frac{\veps^{ab}}{(\veps^{ab})^2 + \gamma^2}
\text{Im} \{
\mathcal{A}_{\nu}^{\pr ab} 
(\mathcal{D}_{\mu} \left[ \mathcal{A}_{\rho}^{\pr} , H \right])^{ba}\}
\\
=&
\frac{2 e^3}{\hbar \gamma} \sum_{\mbk a b} f^a
\frac{\veps^{ab}}{(\veps^{ab})^2 + \gamma^2}
\left[
\text{Im} (C_{\nu\mu\rho}^{ba}) \veps^{ab}
-
\sum_c \text{Re} (S_{\nu\mu\rho}^{abc}) (\veps^{bc} - \veps^{ca})
-
\frac{1}{2} \text{Im} (\mathcal{T}_{\nu\mu\rho}^{ba}) \veps^{ab}
\right],
\end{split}
\end{equation}
\begin{equation}
\begin{split}
\mathcal{F}_3
=&
\frac{2 e^3}{\hbar \gamma^2} \sum_{\mbk a b} f^a
\frac{
\text{Im} \{ \mathcal{A}_{\nu}^{\pr ab} 
\left[ \mathcal{A}_{\mu}^{\pr} , \left[ \mathcal{A}_{\rho}^{\pr} , H \right] \right]^{ba} \}}
{1 + (\eta^{ab})^2}
\\
=&
\frac{2 e^3}{\hbar \gamma^2} \sum_{\mbk a b} f^a
\left[
\text{Re} (\tilde{\mathcal{T}}_{\nu\mu\rho}^{ba}) \veps^{ab}
+
\sum_c \text{Im} (\tilde{Q}_{\nu\mu\rho}^{abc}) \veps^{bc}
-
\sum_c \text{Im} (\tilde{Q}_{\nu\rho\mu}^{abc}) \veps^{ca}
\right],
\end{split}
\end{equation}
\begin{equation}
\begin{split}
\mathcal{F}_{4}
=&
\frac{2 e^3}{\hbar \gamma^2} \sum_{\mbk a} f^a
\text{Im} \big\{
\left( 
\left[
\left[\alpha_{\mu} (\gamma) , H \right]
,
\left[ \mathcal{A}_{\nu}^{\pr} , H \right]
\right] 
\left[ \alpha_{\rho} (-\gamma) , H \right]
\right)_{\mbk}^{a}
\big\}
\\
=&
\frac{2 e^3}{\hbar \gamma^2} \sum_{\mbk abc} f^a
\bigg\{
\text{Im} (Q_{\rho\nu\mu}^{abc})
\frac{ \veps^{ac} \veps^{ab} - \gamma^2}{ [ (\veps^{ac})^2 + \gamma^2 ]
[ (\veps^{ab})^2 + \gamma^2 ] }
-
\text{Re} (Q_{\rho\nu\mu}^{abc})
\frac{ \gamma ( \veps^{ac} + \veps^{ab} ) }{ [ (\veps^{ac})^2 + \gamma^2 ]
[ (\veps^{ab})^2 + \gamma^2 ] }
\\
&-
\text{Im} (Q_{\rho\mu\nu}^{abc})
\frac{ \veps^{cb} \veps^{ab} - \gamma^2}{ [ (\veps^{ab})^2 + \gamma^2 ]
[ (\veps^{bc})^2 + \gamma^2 ] }
+
\text{Re} (Q_{\rho\mu\nu}^{abc})
\frac{ \gamma ( \veps^{cb} + \veps^{ab} ) }{ [ (\veps^{ab})^2 + \gamma^2 ]
[ (\veps^{bc})^2 + \gamma^2 ] }
\bigg\}
\veps^{ab} \veps^{bc} \veps^{ca},
\end{split}
\end{equation}
with $ [ \alpha_{\mu} (\gamma) ]^{ab} \equiv \mathcal{A}_{\mu}^{\pr ab}/( \veps^{ab} - i \gamma)$. Reinserting these back into Eq.~(\ref{pre_sig2_A}) and collecting similar geometric terms, the conductivity reads
\begin{equation}
\label{sig2_A_supp}
\begin{split}
\sigma_{\mu\nu\rho}^{A}
=&
\frac{e^3}{2 \gamma^2}
\sum_{\mbk a} \pd_{\nu} \pd_{\rho} f^a  v_{\mu}^a
-
\frac{e^3}{\hbar^2} \sum_{\mbk ab} f^a
\tilde{\Omega}_{\mu\nu}^{ab} \pd_{\rho} m^{ab}
\frac{2 \eta^{ab}}{1 + (\eta^{ab})^2}
-
\frac{e^3}{\hbar^2} \sum_{\mbk ab} f^a
\tilde{g}_{\mu\nu}^{ab} \pd_{\rho} m^{ab}
\frac{1- (\eta^{ab})^2}{1 + (\eta^{ab})^2}
+
\frac{e^3}{\hbar^2} \sum_{\mbk ab} \pd_{\rho} f^a
\tilde{g}_{\mu\nu}^{ab}  m^{ab}
\\
&-
\frac{e^3}{\hbar^2} \sum_{\mbk ab} f^a
m^{ab} \tilde{\Gamma}_{\nu\rho\mu}^{ab}
+
\frac{e^3}{\hbar^2} \sum_{\mbk ab} f^a
\tilde{g}_{\nu\rho}^{ab} \pd_{\mu} m^{ab}
\frac{1}{(\eta^{ab})^2} 
\frac{1+ 2 (\eta^{ab})^2}{1 + (\eta^{ab})^2}
\\
&-
\frac{e^3}{\hbar^2} \sum_{\mbk ab} f^a m^{ab}
\text{Re} (\tilde{K}_{\mu\nu\rho}^{ba})
-
\frac{e^3}{\hbar^2} \sum_{\mbk ab} f^a m^{ab}
\frac{1}{\eta^{ab}}
\text{Im}
\left(
\tilde{C}_{\mu \nu \rho}^{ba}
-
\tilde{C}_{\nu \mu \rho}^{ba}
-
\tilde{C}_{\rho \nu \mu}^{ba}
\right)
\\
&+
\frac{e^3}{\hbar^2} \sum_{\mbk abc} f^a m^{ab}
\left[
\text{Re} (\tilde{S}_{\mu\nu\rho}^{abc})
- \eta^{ab} \text{Im} (\tilde{S}_{\mu\nu\rho}^{abc})
\right]
\left( \frac{1}{\eta^{bc}} - \frac{1}{\eta^{ca}} \right)
\\
&+
\frac{2 e^3}{\hbar^2} \sum_{\mbk abc} f^a m^{ab}
\text{Re} (\tilde{S}_{\nu\mu\rho}^{abc})
\frac{1}{\eta^{ab}}
\left[
\frac{\eta^{bc}}{\eta^{ca}}
\frac{ 1 + \eta^{ab} \eta^{bc} }{1 + (\eta^{bc})^2}
-
\frac{\eta^{ca}}{\eta^{bc}}
\frac{ 1 + \eta^{ab} \eta^{ca} }{1 + (\eta^{ca})^2}
\right]
\\
&+
\frac{2 e^3}{\hbar^2} \sum_{\mbk abc} f^a m^{ab}
\text{Im} (\tilde{S}_{\nu\mu\rho}^{abc})
\frac{1}{\eta^{ab}}
\left[
\frac{\eta^{bc}}{\eta^{ca}}
\frac{ \eta^{ab} - \eta^{bc} }{1 + (\eta^{bc})^2}
-
\frac{\eta^{ca}}{\eta^{bc}}
\frac{ \eta^{ab} - \eta^{ca} }{1 + (\eta^{ca})^2}
\right]
\\
&-
\frac{2 e^3}{\hbar^2} \sum_{\mbk abc} f^a m^{ab}
\text{Re} (\tilde{A}_{\nu\mu\rho}^{abc})
\frac{1}{\eta^{ab}}
\left[
\frac{1}{\eta^{ca}}
\frac{ \eta^{ab} - \eta^{bc} }{1 + (\eta^{bc})^2}
+
\frac{1}{\eta^{bc}}
\frac{ \eta^{ab} - \eta^{ca} }{1 + (\eta^{ca})^2}
\right]
\\
&+
\frac{2 e^3}{\hbar^2} \sum_{\mbk abc} f^a m^{ab}
\text{Im} (\tilde{A}_{\nu\mu\rho}^{abc})
\frac{1}{\eta^{ab}}
\left[
\frac{\eta^{bc}}{\eta^{ca}}
\frac{ \eta^{ab} - \eta^{bc} }{1 + (\eta^{bc})^2}
+
\frac{\eta^{ca}}{\eta^{bc}}
\frac{ \eta^{ab} - \eta^{ca} }{1 + (\eta^{ca})^2}
\right],
\end{split}
\end{equation}
which reveals the various contributions from the dressed two-state quantum geometric objects. The remaining terms, which cannot be expressed within the two-state formalism, are precisely captured by the symmetric and antisymmetric parts of the three-state QGT. This highlights an interesting--and perhaps useful--parallel with the appearance of the two-state QGT in dissipative linear response theory. As can be seen from Eq.~(\ref{sig_munu_expanded}), the two-state QGT makes an appearance in the response function, which cannot be summed out in general. For example, focussing on the the Berry curvature term responsible for the anomalous Hall effect, 
$\dot{k}^{\nu} \sum_{\mbk ab}
f^a \tilde{\Omega}_{\mu\nu}^{ab}
$,
we note that in the clean limit, one can sum out the intermediate states of the two-state Berry curvature, leaving behind the single-band Berry curvature. That is, in this limit, one could remove all indications that this term is essentially an (intrinsic) interband coherence effect and express it as a single-state object. However, this is no longer the case for the dressed Berry curvature and one should more appropriately consider its band resolution, as a result of the additional interband mixing arising from the dissipation. In this sense, the appearance of the three-state QGT can be regarded as the quadratic-response counterpart to the above argument.

We next repeat the above steps for the distribution derivative terms. This yields the conductivity contribution
\begin{equation}
\begin{split}
\sigma_{\mu\nu\rho}^\text{B}
=&
- \frac{3 e^3}{2 \gamma^2}
\sum_{\mbk a} f^a \pd_{\nu} \pd_{\rho} v_{\mu}^a
-
\frac{e^3}{4 \hbar \gamma} \sum_{\mbk a b} 
\frac{\pd f^a}{\pd \veps^a}
\tilde{\Omega}_{\mu\nu}^{ab} \pd_{\rho} \veps^{ab}
\frac{1+ 5 (\eta^{ab})^2}{1 + (\eta^{ab})^2}
+
\frac{e^3}{\hbar \gamma} \sum_{\mbk a b} 
\pd_{\rho} f^a \tilde{\Omega}_{\mu\nu}^{ab}
+
\frac{e^3}{2\hbar} \sum_{\mbk a b}
\frac{\pd f^a}{\pd \veps^a}
\pd_{\mu} \tilde{g}_{\nu\rho}^{ab}
\\
&+
\frac{e^3}{\hbar \gamma^2} \sum_{\mbk a b}
\frac{\pd f^a}{\pd \veps^a}
\tilde{g}_{\nu \rho}^{ab} \pd_{\mu} \veps^{ab} \veps^{ab}
\frac{1+ 2 (\eta^{ab})^2}{1 + (\eta^{ab})^2}
-
\frac{e^3}{\hbar \gamma^2} \sum_{\mbk a b}
\pd_{\rho}f^a \tilde{g}_{\mu\nu}^{ab} \veps^{ab}
\left[ 1+ 3 (\eta^{ab})^2 \right]
\\
&-
\frac{e^3}{2 \hbar} \sum_{\mbk a b}
\frac{\pd f^a}{\pd \veps^a} \tilde{g}_{\mu\nu}^{ab}
\frac{\pd_{\rho} \veps^{ab}}{\veps^{ab}} 
\frac{1 - 3 (\eta^{ab})^2}{1 + (\eta^{ab})^2}
+
\sum_{i=1}^4 \mathcal{K}_i,
\end{split}
\end{equation}
where
\begin{equation}
\begin{split}
\mathcal{K}_1
=&
\frac{e^3}{2 \hbar \gamma} \sum_{\mbk a b}
\frac{\pd f^a}{\pd \veps^a} \veps^{ab}
\text{Im} \bigg\{
\frac{\mathcal{A}_{\mu}^{\pr ab} 
(\mathcal{D}_{\nu} \left[ \mathcal{A}_{\rho}^{\pr} , H \right])^{ba}}
{\veps^{ab} - i \gamma} \bigg\}
\\
=&
\frac{e^3}{ 2 \hbar \gamma} \sum_{\mbk a b}
\frac{\pd f^a}{\pd \veps^a}
\frac{1}{ 1 + (\eta^{ab})^2 }
\left[
- \frac{1}{2} \Omega_{\mu\rho}^{ab} \pd_{\nu} \veps^{ab}
+
\text{Im} (C_{\mu\nu\rho}^{ba}) \veps^{ab}
-
\sum_c \text{Re} (S_{\mu\nu\rho}^{abc}) (\veps^{bc} - \veps^{ca})
\right]
\\
&+
\frac{e^3}{ 2 \hbar} \sum_{\mbk a b}
\frac{\pd f^a}{\pd \veps^a}
\frac{ \veps^{ab} }{(\veps^{ab})^2 + \gamma^2}
\left[
g_{\mu\rho}^{ab} \pd_{\nu} \veps^{ab}
+
\text{Re} (C_{\mu\nu\rho}^{ba}) \veps^{ab}
+
\sum_c \text{Im} (S_{\mu\nu\rho}^{abc}) (\veps^{bc} - \veps^{ca})
\right],
\end{split}
\end{equation}
\begin{equation}
\begin{split}
\mathcal{K}_2
=&
- \frac{e^3}{\hbar \gamma} \sum_{\mbk a b}
\frac{\pd f^a}{\pd \veps^a}
\frac{\text{Im} \{
\mathcal{A}_{\nu}^{\pr ab} 
(\mathcal{D}_{\mu} \left[ \mathcal{A}_{\rho}^{\pr} , H \right])^{ba}\}}{1 + (\eta^{ab})^2}
\\
=&
- \frac{e^3}{\hbar \gamma} \sum_{\mbk a b}
\frac{\pd f^a}{\pd \veps^a}
\left[
\text{Im} (\tilde{C}_{\nu\mu\rho}^{ba}) \veps^{ab}
-
\sum_c \text{Re} (\tilde{S}_{\nu\mu\rho}^{abc}) (\veps^{bc} - \veps^{ca})
-
\frac{1}{2}
\text{Im} (\tilde{\mathcal{T}}_{\nu\mu\rho}^{ba}) \veps^{ab}
\right],
\end{split}
\end{equation}
\begin{equation}
\begin{split}
\mathcal{K}_3
=&
\frac{ e^3 }{ \hbar } \sum_{\mbk a b}
\frac{\pd f^a}{\pd \veps^a}
\frac{\veps^{ab}}{(\veps^{ab})^2 + \gamma^2}
\text{Im} \{ \mathcal{A}_{\nu}^{\pr ab} 
\left[ \mathcal{A}_{\mu}^{\pr} , \left[ \mathcal{A}_{\rho}^{\pr} , H \right] \right]^{ba} \}
\\
=&
\frac{e^3}{\hbar} \sum_{\mbk a b}
\frac{\pd f^a}{\pd \veps^a}
\left[
\text{Re} (\tilde{\mathcal{T}}_{\nu\mu\rho}^{ba})
+
\sum_c \text{Im} (\tilde{Q}_{\nu\mu\rho}^{abc})
\frac{\veps^{bc}}{\veps^{ab}}
-
\sum_c \text{Im} (\tilde{Q}_{\nu\rho\mu}^{abc})
\frac{\veps^{ca}}{\veps^{ab}}
\right],
\end{split}
\end{equation}
\begin{equation}
\begin{split}
\mathcal{K}_{4}
=&
- \frac{e^3}{\hbar \gamma} \sum_{\mbk a}
\frac{\pd f^a}{\pd \veps^a}
\text{Re} \big\{
\left( 
\left[
\left[\alpha_{\mu} (\gamma) , H \right]
,
\left[ \mathcal{A}_{\nu}^{\pr} , H \right]
\right] 
\left[ \alpha_{\rho} (-\gamma) , H \right]
\right)_{\mbk}^{a}
\big\}
\\
=&
\frac{e^3}{\hbar \gamma} \sum_{\mbk abc}
\frac{\pd f^a}{\pd \veps^a}
\bigg\{
\text{Re} (Q_{\rho\nu\mu}^{abc})
\frac{ \veps^{ac} \veps^{ab} - \gamma^2}{ [ (\veps^{ac})^2 + \gamma^2 ]
[ (\veps^{ab})^2 + \gamma^2 ] }
+
\text{Im} (Q_{\rho\nu\mu}^{abc})
\frac{ \gamma ( \veps^{ac} + \veps^{ab} ) }{ [ (\veps^{ac})^2 + \gamma^2 ]
[ (\veps^{ab})^2 + \gamma^2 ] }
\\
&-
\text{Re} (Q_{\rho\mu\nu}^{abc})
\frac{ \veps^{cb} \veps^{ab} - \gamma^2}{ [ (\veps^{ab})^2 + \gamma^2 ]
[ (\veps^{bc})^2 + \gamma^2 ] }
-
\text{Im} (Q_{\rho\mu\nu}^{abc})
\frac{ \gamma ( \veps^{cb} + \veps^{ab} ) }{ [ (\veps^{ab})^2 + \gamma^2 ]
[ (\veps^{bc})^2 + \gamma^2 ] }
\bigg\}
\veps^{ab} \veps^{bc} \veps^{ca},
\end{split}
\end{equation}
resulting in the contribution
\begin{equation}
\label{sig2_B}
\begin{split}
\sigma_{\mu\nu\rho}^{B}
=&
-\frac{3 e^3}{2 \gamma^2}
\sum_{\mbk a} \pd_{\nu} \pd_{\rho} f^a  v_{\mu}^a
+
\frac{e^3}{\hbar \gamma} \sum_{\mbk ab} \pd_{\rho} f^a
\tilde{\Omega}_{\mu\nu}^{ab}
-
\frac{e^3}{2 \hbar \gamma} \sum_{\mbk ab}
\frac{\pd f^a}{\pd \veps^a}
\tilde{\Omega}_{\mu\nu}^{ab} \pd_{\rho} \veps^{ab}
\frac{1 + 3 (\eta^{ab})^2}{1 + (\eta^{ab})^2}
\\
&+
\frac{e^3}{2 \hbar} \sum_{\mbk ab}
\frac{\pd f^a}{\pd \veps^a}
\pd_{\rho} \tilde{g}_{\mu\nu}^{ab}
-
\frac{3 e^3}{\hbar \gamma} \sum_{\mbk ab}
\pd_{\rho} f^a \tilde{g}_{\mu\nu}^{ab} \eta^{ab}
-
\frac{e^3}{\hbar \gamma^2} \sum_{\mbk ab}
\pd_{\rho} f^a \tilde{g}_{\mu\nu}^{ab} \veps^{ab}
-
\frac{e^3}{\hbar^2} \sum_{\mbk ab}
\frac{\pd f^a}{\pd \veps^a}
\tilde{g}_{\mu\nu}^{ab} \veps^{ab} \pd_{\rho} m^{ab}
\frac{(\eta^{ab})^2}{1 + (\eta^{ab})^2}
\\
&+
\frac{e^3}{2 \hbar} \sum_{\mbk ab}
\frac{\pd f^a}{\pd \veps^a}
\tilde{\Gamma}_{\nu\rho\mu}^{ab}
+
\frac{e^3}{2 \hbar^2} \sum_{\mbk ab}
\frac{\pd f^a}{\pd \veps^a}
\tilde{g}_{\nu\rho}^{ab} \veps^{ab} \pd_{\mu} m^{ab}
\left[ 
\frac{(\eta^{ab})^2}{1 + (\eta^{ab})^2}
-
\frac{2}{(\eta^{ab})^2} 
\frac{1+ 2 (\eta^{ab})^2}{1 + (\eta^{ab})^2}
\right]
\\
&+
\frac{e^3}{2 \hbar} \sum_{\mbk ab}
\frac{\pd f^a}{\pd \veps^a}
\text{Re} (\tilde{K}_{\mu\nu\rho}^{ba})
+
\frac{e^3}{2 \hbar} \sum_{\mbk ab}
\frac{\pd f^a}{\pd \veps^a}
\frac{1}{\eta^{ab}}
\text{Im}
\left(
\tilde{C}_{\mu \nu \rho}^{ba}
-
\tilde{C}_{\nu \mu \rho}^{ba}
-
\tilde{C}_{\rho \nu \mu}^{ba}
\right)
\\
&-
\frac{e^3}{2 \hbar} \sum_{\mbk abc}
\frac{\pd f^a}{\pd \veps^a}
\left[
\text{Re} (\tilde{S}_{\mu\nu\rho}^{abc})
- \eta^{ab} \text{Im} (\tilde{S}_{\mu\nu\rho}^{abc})
\right]
\left( \frac{1}{\eta^{bc}} - \frac{1}{\eta^{ca}} \right)
\\
&+
\frac{e^3}{\hbar} \sum_{\mbk abc}
\frac{\pd f^a}{\pd \veps^a}
\text{Re} (\tilde{S}_{\nu\mu\rho}^{abc})
\left[
\frac{\eta^{bc}}{\eta^{ca}}
\frac{ \eta^{ab} - \eta^{bc} }{1 + (\eta^{bc})^2}
-
\frac{\eta^{ca}}{\eta^{bc}}
\frac{ \eta^{ab} - \eta^{ca} }{1 + (\eta^{ca})^2}
\right]
\\
&-
\frac{e^3}{\hbar} \sum_{\mbk abc}
\frac{\pd f^a}{\pd \veps^a}
\text{Im} (\tilde{S}_{\nu\mu\rho}^{abc})
\left[
\frac{\eta^{bc}}{\eta^{ca}}
\frac{ 1 + \eta^{ab} \eta^{bc} }{1 + (\eta^{bc})^2}
-
\frac{\eta^{ca}}{\eta^{bc}}
\frac{ 1 + \eta^{ab} \eta^{ca} }{1 + (\eta^{ca})^2}
\right]
\\
&+
\frac{e^3}{\hbar} \sum_{\mbk abc}
\frac{\pd f^a}{\pd \veps^a}
\text{Re} (\tilde{A}_{\nu\mu\rho}^{abc})
\left[
\frac{1}{\eta^{ca}}
\frac{ 1 + \eta^{ab} \eta^{bc} }{1 + (\eta^{bc})^2}
+
\frac{1}{\eta^{bc}}
\frac{ 1 + \eta^{ab} \eta^{ca} }{1 + (\eta^{ca})^2}
\right]
\\
&-
\frac{e^3}{\hbar} \sum_{\mbk abc}
\frac{\pd f^a}{\pd \veps^a}
\text{Im} (\tilde{A}_{\nu\mu\rho}^{abc})
\left[
\frac{\eta^{bc}}{\eta^{ca}}
\frac{ 1 + \eta^{ab} \eta^{bc} }{1 + (\eta^{bc})^2}
+
\frac{\eta^{ca}}{\eta^{bc}}
\frac{ 1 + \eta^{ab} \eta^{ca} }{1 + (\eta^{ca})^2}
\right].
\end{split}
\end{equation}

Adding all the various conductivity contributions, the total current density can be obtained. To do so, we note that the distribution derivative terms in Eq.~(\ref{sig2_B}) can be reexpressed with the identity
\begin{equation}
\sum_{\mbk} \frac{\pd f^a}{\pd \veps^a} \mathcal{O}
=
- \sum_{\mbk} f^a \pd^{\sigma}
\left[
\frac{1}{\hbar} \frac{v_{\sigma}^a}{(\bs{v}^a)^2} \mathcal{O}
\right]
=
- \sum_{\mbk} f^a
\bigg\{ \pd^{\sigma} , 
\frac{1}{\hbar} \frac{v_{\sigma}^a}{(\bs{v}^a)^2} \bigg\}\mathcal{O},
\end{equation}
where $\mathcal{O}$ is an arbitrary momentum- and band-dependent function and the curly brackets in the last term denote the anticommutator.

Collecting terms from the linear and nonlinear conductivities, we ultimately arrive at the total current density
\begin{equation}
\label{j_mu_supp}
\begin{split}
\frac{j_{\mu}}{-e}
=&
\sum_{\mbk a} (n_1^a + n_2^a) v_{\mu}^a
+
\dot{k}^{\nu} \sum_{\mbk ab} (f^a + n_1^a)
\mathcal{Z}_{\Omega}^{ab} \tilde{\Omega}_{\mu \nu}^{ab}
-
\dot{k}^{\nu} \sum_{\mbk ab} (f^a + n_1^a)
\mathcal{Z}_{g}^{ab} \tilde{g}_{\mu \nu}^{ab}
+
\dot{k}^{\nu} \dot{k}^{\rho}
\sum_{\mbk ab} f^a m^{ab}
\mathcal{Z}_{\Gamma}^{ab}
\tilde{\Gamma}_{\nu \rho \mu}^{ba}
\\
&+
\dot{k}^{\nu} \dot{k}^{\rho} \sum_{\mbk ab} f^a 
\mathcal{Z}_{m}^{ab} 
\tilde{g}_{\nu \rho}^{ab} \pd_{\mu} m^{ab}
+
\dot{k}^{\nu} \dot{k}^{\rho}
\sum_{\mbk ab} f^a m^{ab}
\left[
\mathcal{Z}_{K}^{ab}
\text{Re} (\tilde{\mathcal{K}}_{\mu\nu\rho}^{ba})
+
\mathcal{Z}_{C}^{ab}
\text{Im}
\left(
\tilde{C}_{\mu \nu \rho}^{ba}
-
\tilde{C}_{\nu \mu \rho}^{ba}
-
\tilde{C}_{\rho \nu \mu}^{ba}
\right)
\right]
\\
&+
\dot{k}^{\nu} \dot{k}^{\rho}
\sum_{\mbk abc} f^a m^{ab}
\left[
\mathcal{Z}_{1}^{abc}
\text{Re} (\tilde{S}_{\mu\nu\rho}^{abc})
+
\mathcal{Z}_{2}^{abc}
\text{Im} (\tilde{S}_{\mu\nu\rho}^{abc})
+
\mathcal{Z}_{3}^{abc}
\text{Re} (\tilde{S}_{\nu\mu\rho}^{abc})
+
\mathcal{Z}_{4}^{abc}
\text{Im} (\tilde{S}_{\nu\mu\rho}^{abc})
\right]
\\
&+
\dot{k}^{\nu} \dot{k}^{\rho}
\sum_{\mbk abc} f^a m^{ab}
\left[
\mathcal{Z}_{5}^{abc}
\text{Re} (\tilde{A}_{\nu\mu\rho}^{abc})
+
\mathcal{Z}_{6}^{abc}
\text{Im} (\tilde{A}_{\nu\mu\rho}^{abc})
\right],
\end{split}
\end{equation}
where $n_2^a = - \tau \dot{k}^{\nu} \pd_{\nu} n_1^a$ yields the nonlinear Drude weight and the various renormalization functions that appear are given by
\begin{equation}
\mathcal{Z}_{\Omega}^{ab}
=
1 + \dot{k}^{\rho} \pd_{\rho} m^{ab}
\frac{2 \eta^{ab}}{1 + (\eta^{ab})^2}
+
\frac{1}{2} \bigg\{ \pd^{\sigma} , 
\frac{v_{\sigma}^a}{(\bs{v}^a)^2}
\dot{k}^{\rho} \pd_{\rho} \ln (m^{ab})
\frac{1}{\eta^{ab}} 
\frac{1 + 3 (\eta^{ab})^2}{1 + (\eta^{ab})^2}
\bigg\},
\end{equation}
\begin{equation}
\begin{split}
\mathcal{Z}_{g}^{ab}
=&
2 \eta^{ab} - \dot{k}^{\rho} \pd_{\rho} m^{ab}
\frac{1 -  (\eta^{ab})^2 }{ 1 + (\eta^{ab})^2}
-
\frac{1}{2} \bigg\{ 
\pd^{\sigma} , 
\frac{v_{\sigma}^a}{(\bs{v}^a)^2}
\bigg\}
\dot{k}^{\rho} \pd_{\rho}
+
\frac{1}{(\eta^{ab})^2} 
\left[
m^{ab} , \dot{k}^{\rho} \pd_{\rho}
\right]
+
\bigg\{ \pd^{\sigma} , 
\frac{v_{\sigma}^a}{(\bs{v}^a)^2}
\dot{k}^{\rho} \pd_{\rho} \ln (m^{ab})
\frac{(\eta^{ab})^2}{1 + (\eta^{ab})^2}
\bigg\},
\end{split}
\end{equation}
\begin{equation}
\mathcal{Z}_{m}^{ab}
=
- \frac{1}{(\eta^{ab})^2}
\frac{1+ 2 (\eta^{ab})^2}{1 + (\eta^{ab})^2}
+
\frac{1}{2} \bigg\{ 
\pd^{\sigma} ,
\frac{1}{m^{ab}}
\frac{v_{\sigma}^a}{(\bs{v}^a)^2}
\left[
\frac{(\eta^{ab})^2}{1 + (\eta^{ab})^2}
-
\frac{2}{(\eta^{ab})^2}
\frac{ 1 + 2 (\eta^{ab})^2}{1 + (\eta^{ab})^2}
\right]
\bigg\},
\end{equation}
\begin{equation}
\mathcal{Z}_{\Gamma}^{ab}
=
\mathcal{Z}_{K}^{ab}
=
1
+
\frac{1}{2 m^{ab}} \bigg\{ 
\pd^{\sigma} , 
\frac{v_{\sigma}^a}{(\bs{v}^a)^2}
\bigg\}
\end{equation}
\begin{equation}
\mathcal{Z}_{C}^{ab}
=
\frac{1}{\eta^{ab}}
+
\frac{1}{2 m^{ab}} \bigg\{ 
\pd^{\sigma} , 
\frac{v_{\sigma}^a}{(\bs{v}^a)^2}
\frac{1}{\eta^{ab}}
\bigg\}
\end{equation}
\begin{equation}
\mathcal{Z}_{1}^{abc}
=
\frac{1}{\eta^{ca}}
-
\frac{1}{\eta^{bc}}
+
\frac{1}{2 m^{ab}} \bigg\{ 
\pd^{\sigma} , 
\frac{v_{\sigma}^a}{(\bs{v}^a)^2}
\left(
\frac{1}{\eta^{ca}}
-
\frac{1}{\eta^{bc}}
\right)
\bigg\},
\end{equation}
\begin{equation}
\mathcal{Z}_{2}^{abc}
=
\frac{\eta^{ab}}{\eta^{bc}}
-
\frac{\eta^{ab}}{\eta^{ca}}
+
\frac{1}{2 m^{ab}} \bigg\{ 
\pd^{\sigma} , 
\frac{v_{\sigma}^a}{(\bs{v}^a)^2}
\left(
\frac{\eta^{ab}}{\eta^{bc}}
-
\frac{\eta^{ab}}{\eta^{ca}}
\right)
\bigg\},
\end{equation}
\begin{equation}
\mathcal{Z}_{3}^{abc}
=
\frac{2}{\eta^{ab}}
\left[
\frac{\eta^{ca}}{\eta^{bc}}
\frac{1+\eta^{ab}\eta^{ca}}{1+(\eta^{ca})^2}
-
\frac{\eta^{bc}}{\eta^{ca}}
\frac{1+\eta^{ab}\eta^{bc}}{1+(\eta^{bc})^2}
\right]
-
\frac{1}{m^{ab}} \bigg\{ 
\pd^{\sigma} , 
\frac{v_{\sigma}^a}{(\bs{v}^a)^2}
\left[
\frac{\eta^{ca}}{\eta^{bc}}
\frac{\eta^{ab} - \eta^{ca}}{1+(\eta^{ca})^2}
-
\frac{\eta^{bc}}{\eta^{ca}}
\frac{\eta^{ab} - \eta^{bc}}{1+(\eta^{bc})^2}
\right]
\bigg\},
\end{equation}
\begin{equation}
\mathcal{Z}_{4}^{abc}
=
\frac{2}{\eta^{ab}}
\left[
\frac{\eta^{ca}}{\eta^{bc}}
\frac{\eta^{ab} - \eta^{ca}}{1+(\eta^{ca})^2}
-
\frac{\eta^{bc}}{\eta^{ca}}
\frac{\eta^{ab} - \eta^{bc}}{1+(\eta^{bc})^2}
\right]
+
\frac{1}{m^{ab}} \bigg\{ 
\pd^{\sigma} , 
\frac{v_{\sigma}^a}{(\bs{v}^a)^2}
\left[
\frac{\eta^{ca}}{\eta^{bc}}
\frac{1+\eta^{ab}\eta^{ca}}{1+(\eta^{ca})^2}
-
\frac{\eta^{bc}}{\eta^{ca}}
\frac{1+\eta^{ab}\eta^{bc}}{1+(\eta^{bc})^2}
\right]
\bigg\},
\end{equation}
\begin{equation}
\mathcal{Z}_{5}^{abc}
=
\frac{2}{\eta^{ab}}
\left[
\frac{1}{\eta^{ca}}
\frac{\eta^{ab} - \eta^{bc}}{1+(\eta^{bc})^2}
+
\frac{1}{\eta^{bc}}
\frac{\eta^{ab} - \eta^{ca}}{1+(\eta^{ca})^2}
\right]
+
\frac{1}{m^{ab}} \bigg\{ 
\pd^{\sigma} , 
\frac{v_{\sigma}^a}{(\bs{v}^a)^2}
\left[
\frac{1}{\eta^{ca}}
\frac{1 + \eta^{ab}\eta^{bc}}{1+(\eta^{bc})^2}
+
\frac{1}{\eta^{bc}}
\frac{1 + \eta^{ab}\eta^{ca}}{1+(\eta^{ca})^2}
\right]
\bigg\},
\end{equation}
\begin{equation}
\mathcal{Z}_{6}^{abc}
=
- \frac{2}{\eta^{ab}}
\left[
\frac{\eta^{bc}}{\eta^{ca}}
\frac{\eta^{ab} - \eta^{bc}}{1+(\eta^{bc})^2}
+
\frac{\eta^{ca}}{\eta^{bc}}
\frac{\eta^{ab} - \eta^{ca}}{1+(\eta^{ca})^2}
\right]
+
\frac{1}{m^{ab}} \bigg\{ 
\pd^{\sigma} , 
\frac{v_{\sigma}^a}{(\bs{v}^a)^2}
\left[
\frac{\eta^{bc}}{\eta^{ca}}
\frac{1 + \eta^{ab}\eta^{bc}}{1+(\eta^{bc})^2}
+
\frac{\eta^{ca}}{\eta^{bc}}
\frac{1 + \eta^{ab}\eta^{ca}}{1+(\eta^{ca})^2}
\right]
\bigg\}.
\end{equation}
From these, we observe that the distribution derivative terms in the response functions not only affect the carrier densities, but can also be thought of as inducing gradient corrections to the various geometric structures that appear in the equation of motion through the renormalization functions. Comparing Eq.~(\ref{j_mu_supp}) with the generic relation for the current density, $\sum_{\mbk a} n^a \dot{x}_{\mu}^a$, where $n^a = f^a + n_1^a + n_2^a$, the dressed equation of motion of the carrier position, Eq.~(9) in the main text, is obtained.
 
\section{S2.~Derivation of Dissipative field equations}

To derive the dissipative field equations, Eq.~(14) in the main text, it is helpful to recall the expression for the Levi-Civita connection of an arbitrary rank $(p,q)$ tensor $\mathcal{O}$~\cite{carroll2019spacetime}
\begin{equation}
\begin{split}
\nabla_{\rho} \tensor{\mathcal{O}}{^{\mu_1}^{\cdots}^{\mu_p}_{\nu_1}_{\cdots}_{\nu_q}}
&=
\pd_{\rho} \tensor{\mathcal{O}}{^{\mu_1}^{\cdots}^{\mu_p}_{\nu_1}_{\cdots}_{\nu_q}}
+
\Gamma_{\rho \sigma}^{\mu_1}
\tensor{\mathcal{O}}{^{\sigma}^{\cdots}^{\mu_p}_{\nu_1}_{\cdots}_{\nu_q}}
+
\cdots
+
\Gamma_{\rho \sigma}^{\mu_p}
\tensor{\mathcal{O}}{^{\mu_1}^{\cdots}^{\sigma}_{\nu_1}_{\cdots}_{\nu_q}}
\\
&-
\Gamma_{\rho \nu_1}^{\sigma}
\tensor{\mathcal{O}}{^{\mu_1}^{\cdots}^{\mu_p}_{\sigma}_{\cdots}_{\nu_q}}
-
\cdots
-
\Gamma_{\rho \nu_q}^{\sigma}
\tensor{\mathcal{O}}{^{\mu_1}^{\cdots}^{\mu_p}_{\nu_1}_{\cdots}_{\sigma}},
\end{split}
\end{equation}
as well as the Riemann tensor in terms of connection components
\begin{equation}
\tensor{R}{^{\mu}_\nu_\rho_\sigma}
=
\pd_{\rho} \Gamma^{\mu}_{\nu \sigma}
-
\pd_{\sigma} \Gamma^{\mu}_{\nu \rho}
+
\Gamma^{\lambda}_{\nu \sigma}
\Gamma^{\mu}_{\lambda \rho}
-
\Gamma^{\lambda}_{\nu \rho}
\Gamma^{\mu}_{\lambda \sigma}.
\end{equation}
Following the procedure to obtain dressed quantities as discussed in the main text, we perform the necessary contractions to obtain the dressed Ricci tensor and scalar as
\begin{equation}
\tilde{R}_{\mu\nu}
=
R_{\mu\nu} - \frac{1}{2} g_{\mu\nu} \nabla^2 \lambda
-
\frac{n-2}{2} \nabla_{\mu} \nabla_{\nu}\lambda,
\end{equation}
and
\begin{equation}
\tilde{R}
=
\lambda^{-1}R
-
(n+1) \nabla^2 \lambda.
\end{equation}
Combining terms results in the dissipative contributions to the Einstein tensor, $\tilde{G}_{\mu\nu} - G_{\mu\nu}$, from which the field equations are obtained.

\end{document}